\documentclass[11pt,twoside]{book} 
\usepackage{asp2008s}
\usepackage{times}
\usepackage{lscape}
\usepackage{epsf}

\usepackage{graphicx}
\usepackage{amssymb}
\usepackage{longtable}
\usepackage[figuresright]{rotating}
\usepackage{multirow}

\newcommand{\CO}{$^{12}${\rmfamily CO}{(2--1)}$\,$}

\newcommand{\CII}{\ion{C}{ii}}

\newcommand{\HII}{$\mathrm{H\,{\scriptstyle II}}\,$}

\newcommand{\kms}{km~s{$^{-1}$}}

\newcommand{\Msun}{M{$_{\odot}$}}

\newcommand{\OI}{\ion{O}{i}}

\newcommand{\simless}{\mathbin{\lower 3pt\hbox {$\rlap{\raise 5pt\hbox{$\char'074$}}\mathchar"7218$}}}

\mainmatter

\newlength{\deftabcolsep}
\begin{document}

\title{The Carina Nebula: A Laboratory for Feedback and Triggered Star
Formation}

\author{Nathan Smith}
\affil{Astronomy Department, University of California, 601 Campbell
  Hall, Berkeley, CA 94720, USA}

\author{Kate J.\ Brooks}
\affil{ATNF, P.O. Box 76, Epping NSW 1710, Australia}

\begin{abstract}
The Carina Nebula (NGC~3372) is our richest nearby laboratory in which
to study feedback through UV radiation and stellar winds from very
massive stars during the formation of an OB association, at an early
phase before supernova explosions have disrupted the environment.
This feedback is triggering new generations of star formation around
the periphery of the nebula, while simultaneously evaporating the gas
and dust reservoirs out of which young stars are trying to accrete.
Carina is currently powered by UV radiation from 65 O-type stars and 3
WNH stars, but for most of its lifetime when its most massive star
($\eta$ Carinae) was on the main-sequence, the Carina Nebula was
powered by 70 O-type stars that produced a hydrogen ionizing
luminosity 150 times stronger than in the Orion Nebula.  At a distance
of 2.3 kpc, Carina has the most extreme stellar population within a
few kpc of the Sun, and suffers little interstellar extinction.  It is
our best bridge between the detailed star-formation processes that can
be studied in nearby regions like Orion, and much more extreme but
also more distant regions like 30 Doradus.  Existing observations have
only begun to tap the tremendous potential of this region for
understanding the importance of feedback in star formation --- it will
provide a reservoir of new discoveries for the next generation of
large ground-based telescopes, space telescopes, and large
submillimeter and radio arrays.
\end{abstract}

\section{Introduction}

Feedback from young massive stars may play an integral role in star
and planet formation.  Most stars are born in OB associations (Blaauw
1964; Lada \& Lada 2003), in the vicinity of hot massive stars spawned
only from giant molecular clouds.  The effects of feedback from these
massive stars cannot be studied in nearby quiescent regions of star
formation.

\begin{figure}[!ht]
\centering
\includegraphics[draft=False,width=0.95\textwidth]{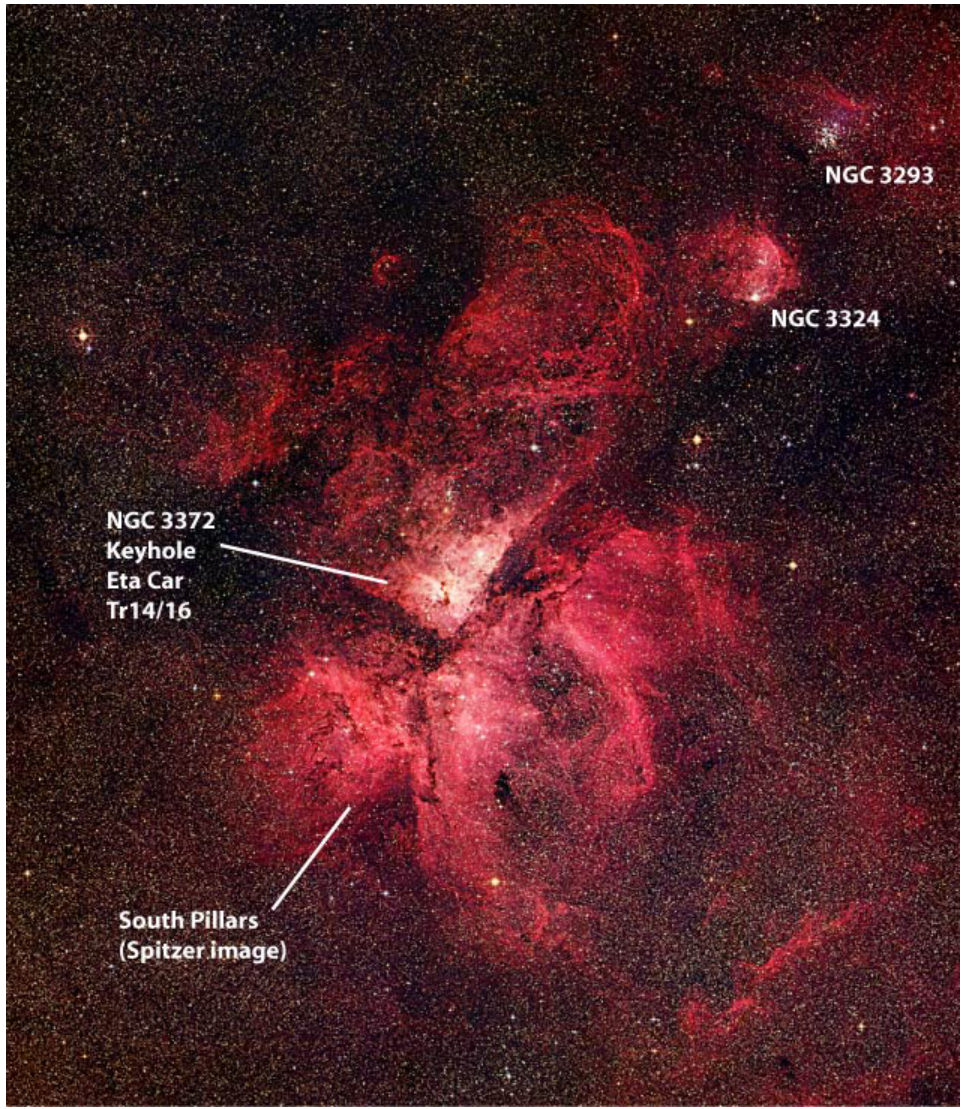}
\caption{A large field-of-view photograph of the Carina Nebula made
  from UK Schmidt plates by David Malin.
  Anglo-Australian Observatory/Royal Obs.\ Edinburgh.}
\label{fig:big}
\end{figure}

The Carina Nebula (Fig.~\ref{fig:big}; NGC~3372) is the southern
hemisphere's largest and highest surface brightness nebula --- some
central parts of the nebula are even brighter than the Orion Nebula.
It provides an ideal laboratory in which to study ongoing star
formation in the vicinity of some of the most massive stars known,
including our Galaxy's most luminous known star, $\eta$ Carinae.
Carina's star clusters are not as densely packed as those of 30~Dor in
the LMC (Massey \& Hunter 1998), or objects in our own Galaxy like the
Arches cluster near the Galactic center (Najarro et al.\ 2004; Figer
et al.\ 1999, 2002), NGC~3603 (Moffat et al.\ 2002), or W49 (Welch et
al.\ 1987).  However, these other regions are too distant for detailed
studies of small-scale phenomena like irradiated protoplanetary disks
and jets, and their study is hampered by considerably more extinction.
The low extinction toward Carina combined with its proximity and rich
nebular content provide a worthwhile trade-off.

For the most massive stars, lifetimes of $\sim$3 Myr before they
explode as supernovae (SNe) are comparable to the time it takes to
clear away their surrounding large-scale distribution of molecular
material.  Consequently, we already have a situation in Carina where
the most massive members like $\eta$~Car are approaching their
imminent demise while new stars are being born from dense molecular
gas only 5--20 pc away.  In the next 1--2 Myr, there will be several
energetic SNe in Carina.  These will carve out an even larger cavity
in the ISM and form a giant superbubble in the Galactic plane, and may
pollute protoplanetary disks with nuclear-processed ejecta.  In the
mean time, studying this rich region provides a snapshot of a young
proto-superbubble (Smith et al.\ 2000) energized only by UV radiation
and stellar winds, just before the disruptive SN-dominated phase.

\begin{figure}[!ht]
\centering
\includegraphics[draft=False,width=0.95\textwidth]{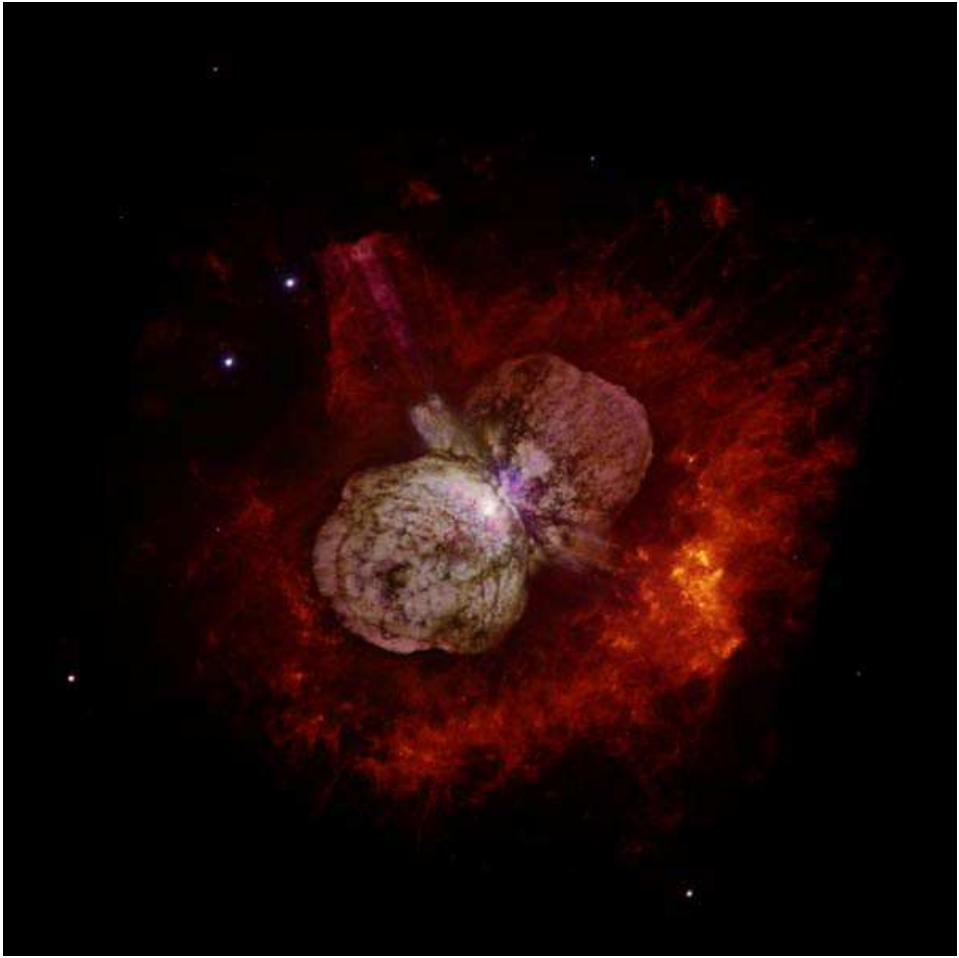}
\caption{An {\it HST}/WFPC2 image of $\eta$ Carinae and its
  surrounding nebulosity (N.~Smith/NASA).}
\label{fig:eta1}
\end{figure}
\begin{figure}[!ht]
\centering
\includegraphics[draft=False,width=0.8\textwidth]{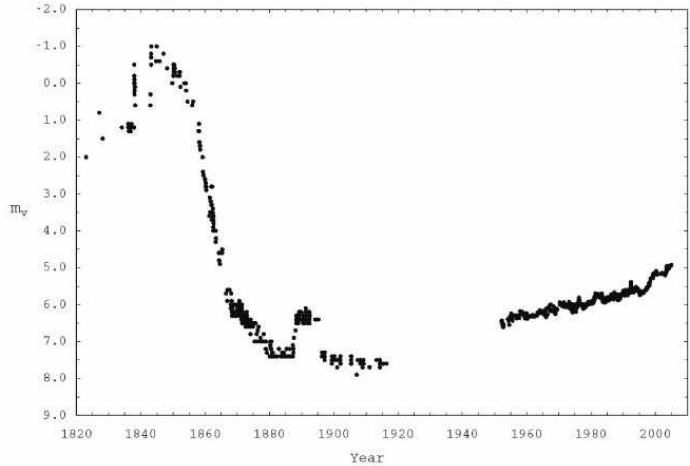}
\caption{The historical visual light curve of $\eta$ Carinae (from
  Frew 2004).}
\label{fig:lightcurve}
\end{figure}

\subsection{The Star:  Eta Carinae}

From before the time of Herschel up to the present, most observational
effort in the Carina Nebula had focussed on the variable star $\eta$
Carinae (Fig.~\ref{fig:eta1}) and its immediate surroundings.  The
unusual variability of $\eta$~Car was noted as early as the 17th
century by Halley, when it was reputed to vary between 4th and 2nd
magnitude.  Later, in the early and mid 19th century, it was observed
by John Herschel (1847) as it brightened significantly to become the
second brightest star in the sky.  It then faded from view by the
1870s.  Frew (2004) gives a careful and detailed account of the
historical light curve (Fig.~\ref{fig:lightcurve}).  During that
event, the star ejected more than 10~$M_{\odot}$ with almost 10$^{50}$
erg of kinetic energy (Smith et al.\ 2003b), which has since expanded
to form the so-called ``Homunculus'' nebula seen today in {\it HST}
images like the one in Figure~\ref{fig:eta1}.  After more than 160
years and intensive study at all wavelengths, the underlying cause of
this so-called ``Great Eruption'' is still the most enduring mystery
associated with $\eta$~Car.  Additionally, much current work centers
around the 5.5-yr periodic variability that is likely the product of
$\eta$ Carinae being an eccentric binary system (e.g., Damineli et
al.\ 2000).

Our focus here is on star formation in the surrounding giant H~{\sc
ii} region, so we do not aim to review or summarize the vast and
complicated literature concerning $\eta$ Carinae and its ejecta (see
Davidson \& Humphreys 1997; although much has been added to our
understanding in the subsequent decade making that review outdated in
some respects).  However, the wild variability of $\eta$~Car, its
extreme nature, and its imminent demise are relevant for nebular
structures in the region.  From the point of view of understanding
star formation in the region, two important facts are most worth
remembering:

(1) Although $\eta$ Carinae is probably the most massive and luminous
    star known in the Milky Way, {\it at the current time it
    contributes essentially nothing to the radiative energy budget of
    the surrounding region}.  This strange twist arises because $\eta$
    Carinae is currently surrounded by a nearly opaque dust shell
    ejected 160 years ago, which absorbs nearly all of the star's UV
    and visual luminosity.  That luminosity is then re-radiated in the
    thermal IR, and it escapes from the nebula, exerting little
    influence on the surrounding gas in the star-forming complex.

(2) However, before 1843 (for about 3 Myr prior), $\eta$~Carinae had a
    tremendous influence on the UV energy budget of the entire region.
    Its UV luminosity dominated the ionization of the main cavity and
    it sculpted many of the most prominent elephant trunks in the
    region.  When the giant eruption ended about 150 years ago and
    $\eta$ Car faded due to dust obscuration, the UV luminosity of the
    region dropped by 20\% (Smith 2006a) and the remaining O stars
    that are scattered throughout the region now dominate the
    ionization.  This change, which arises because of $\eta$ Car's
    precarious instability and its evolved state, is unique among
    Galactic H~{\sc ii} regions.

\begin{figure}[!h]
\centering
\includegraphics[draft=False,width=0.95\textwidth]{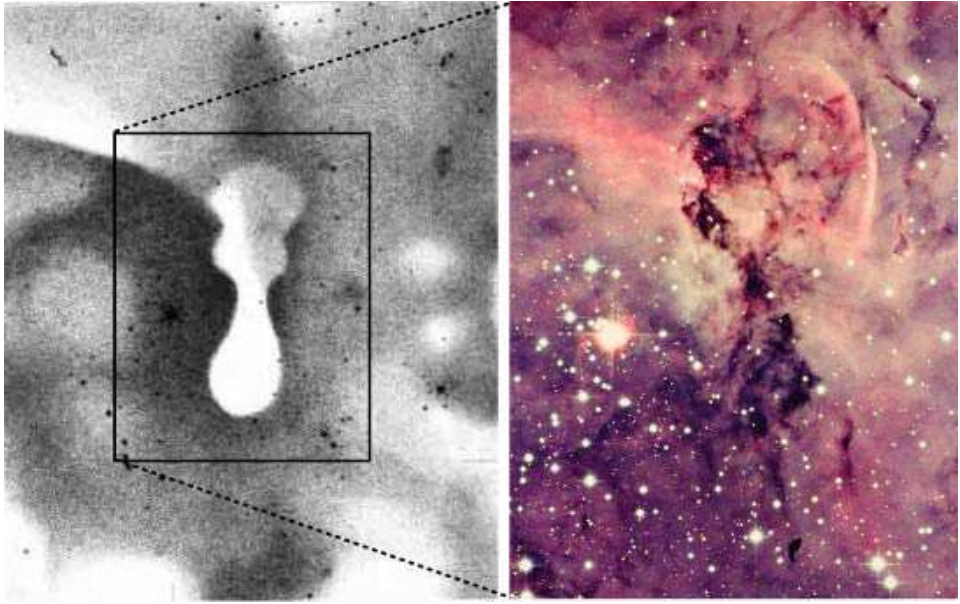}
\caption{Left: Herschel's drawing of the Keyhole Nebula, made during
  $\eta$ Carinae's 19th century outburst.  Right: A modern color
  photograph of the Keyhole made by David Malin.}
\label{fig:herschel}
\end{figure}

\subsection{Historical Perspective: The Keyhole}

The famous ``Keyhole'' nebula at the heart of the greater Carina
Nebula also has a rich history, aided by its location immediately
adjacent to $\eta$ Carinae, as well as its extremely high surface
brightness.  The physical origin of the Keyhole shape is still not
understood (although see Smith 2002b).

While the old-fashioned keyhole shape is easily recognized in
Herschel's drawing, more creativity is required to identify this
structure in modern photographs (see comparison in
Fig.~\ref{fig:herschel}).  The most dramatic difference is in the
lower part of the Keyhole, where much of the diffuse emission is
brighter in Herschel's drawing.  This has caused some consternation to
observers (e.g., Bok 1932), making some suspect that Herschel may have
had an over-active imagination.  In fact, it is likely that Herschel's
drawing was largely correct, and that the appearance of the nebula has
actually changed with time.  Herschel's drawing in
Figure~\ref{fig:herschel} was made around 1840, during the Great
Eruption of $\eta$~Car, before the star was obscured by its dusty
nebula.  Thus, at that time the Keyhole was illuminated by a much
brighter star than is seen today, and the reflection nebulosity around
the Keyhole could have differed in appearance.  We now know from both
historical and modern data that the reflected spectrum of $\eta$~Car
is seen in the Keyhole (Walborn \& Liller 1977; Lopez \& Meaburn
1986), confirming that it is at the same distance as $\eta$~Car.
Given that the larger Carina Nebula is more than 150 light years
across, one might wonder if light echoes from the Great Eruption can
still be seen in archival images.

As for the diffuse emission from the larger Carina Nebula, Herschel
did have a few choice words to describe it: ``It would be manifestly
impossible by verbal description to give any just idea of the
capricious forms and irregular gradations of light affected by the
different branches and appendages of this nebula.''  The rich, complex
structure that Herschel was referring to is associated with the many
dust pillars and dark globules, the likes of which are not seen in the
Orion Nebula.  Perhaps this is an early testament to feedback from the
extreme stellar population in Carina, providing us with a snapshot of
a molecular cloud getting shredded by its progeny.

In the 20th century, observers began to take a broader view of Carina,
realizing that it spanned several degrees of the sky.  In this review,
we are concerned with star formation across the full extent of the
Carina Nebula.  The nomenclature for this region is interchanged
loosely, being referred to variously as NGC~3372, the ``Carina
Nebula'', the ``Great Carina Nebula'', the ``Keyhole Nebula'', the
``Eta Carinae Nebula'', or (incorrectly) as simply ``Eta Carinae''.
We reserve ``Eta Carinae'' for discussions of the massive star itself,
the ``Homunculus'' nebula (Gaviola 1950), plus surrounding ejecta
nebulosity within about 1\arcmin\ (e.g.\ Smith et al.\ 2005a).  We use
the name ``Keyhole Nebula'' to refer to the structures in
Figure~\ref{fig:herschel}, within about 10\arcmin\ of $\eta$ Car,
while we use the term ``Carina Nebula'' to refer to the larger giant
H~{\sc ii} region spanning several square degrees (Fig.~\ref{fig:big}),
of which $\eta$ Car and the Keyhole are only a small component at the
center.

\subsection{Pioneering Observations of the Nebulosity}

The first modern interference-filter photographs of the region were
made by Walborn (1975) and Deharveng \& Maucherat (1975).  These
images showed the complex structures of the Keyhole and the rest of
NGC~3372 in ionized gas, and revealed many silhouette structures, such
as dust pillars and globules, and the large, dark V-shaped dust lane
that bisects the nebula.  It was established early on that the
V-shaped dark lanes are the result of obscuration by dust intermixed
with molecular gas (\citealt{Dickel741}; \citealt{Feinstein73};
\citealt{Smith87}).  Far-infrared (far-IR) emission was detected from
the dark lanes by \citet{Harvey79} at 80~$\mu$m and \citet{Ghosh88} at
120--300 $\mu$m, implying that the dust was warm, with T$\simeq$30--40
K.  The first single-dish molecular line studies of Carina by Gardner
et al.\ (1973) and \citet{Dickel742} revealed that these dark dust
lanes coincided with the bulk of the molecular gas.

Pioneering radio observations of the Carina nebula focused on the
central region with angular resolutions (FWHM) that were typically
3\arcmin\ (e.g., \citealt{Gardner68}; \citealt{Shaver70};
\citealt{Gardner70}; \citealt{Jones73}; \citealt{Dickel741};
\citealt{Huchtmeier75}; \citealt{Retallack83};
\citealt{Tateyama91}). \citet{Gardner70} first identified the two
bright concentrations known as Car~I and Car~II (see
Fig.~\ref{fig:co}).  Subsequent observations established that these
two sources were thermal and represented two separate H~{\sc ii}
regions. Car I was found to be located toward the north-western part
of the inner nebula, and Car II was identified with the central
Keyhole Nebula. \citet{Retallack83} was the first to separate
$\eta$~Car from the bright radio continuum emission of Car~II, and to
highlight the similar ring-shaped structures of the radio source
Car~II and the Keyhole Nebula seen at visual wavelengths.

Detections of non-thermal emission toward the nebula were reported by
\citet{Jones73} and \citet{Tateyama91}. However, no subsequent work
has found convincing evidence to confirm these non-thermal sources.
The non-thermal source G287.7--1.3 first reported by \citet{Shaver70}
has since been shown to be extragalactic \citep{Caswell75}.  The
magnetar studied by Gaensler et al.\ (2005) is located at a larger
distance and well outside the Carina Nebula 1$^{\circ}$ to the east.

It is amusing to note that the first large-scale CO study of the
Carina Nebula at 115 GHz used the {\it optical} ESO 3.6-m telescope at
La Silla in 1978 \citep{deGraauw81}. These data were soon followed by
the CO survey undertaken by \citet{ Whiteoak84} using the 4-m radio
telescope in Sydney.  The resulting maps with a $\sim$2\arcmin\ (FWHM)
beam offered the most detailed view of the Carina molecular cloud for
more than a decade, and showed that the bulk of the molecular material
was concentrated into a northern and a southern cloud.  The total
extent of the molecular cloud was revealed later in the Columbia CO
survey of the Galactic plane \citep{Grabelsky88}. The survey showed
that the Carina molecular cloud is part of a GMC complex that is
situated on the near side of the Carina arm and extends over 150 pc
between $l$=284.7 and 289, with a total estimated mass of
6.7$\times$10$^5$~M$_{\odot}$.

Results from the first studies of the velocity field within the Carina
Nebula showed complex structure that is still not
understood. Interpretations of the large-scale dynamics measured
across the nebula involved merging spiral arms \citep{Tateyama91},
rotating neutral clouds \citep{Meaburn84} or old \HII\ regions
\citep{Cersosimo84}. On a smaller scale, the complicated dynamics
measured in the vicinity of Car~II alluded to the tremendous influence
of the stellar wind from $\eta$~Car. Motions of ionized gas were
studied via radio hydrogen recombination lines
(e.g. \citealt{Gardner70}; \citealt{Huchtmeier75};
\citealt{Cersosimo84}) and optical-line emission profiles
(e.g. \citealt{Deharveng75}). All studies detected double-peaked
profiles in the vicinity of Car II. \citet{Deharveng75} proposed that
the ionized gas lies in an expanding shell whose center of expansion
is $\eta$~Car.

Additional evidence for peculiar, high-speed gas motions in the
vicinity of $\eta$ Car has been seen in resonance-line absorption
profiles (\citealt{Walborn75}).  These unusual multi-component
profiles have received continued study in the optical and UV (Walborn
1982; Walborn \& Hesser 1982; \citealt{Laurent82}; Garcia \& Walborn
2000; Walborn et al.\ 2002a, 2007), but are still not understood.
Velocity components range over 500 km s$^{-1}$ and have sometimes been
interepreted as evidence for a supernova remnant along the line of
sight, although proof of this hypothesis remains elusive.  Elliot
(1979) interpreted broad emission components as evidence for a
supernova remnant in the Keyhole, but these were actually broad lines
in the wind spectrum of $\eta$ Car itself that were reflected by dust
in the Keyhole (see Walborn \& Liller 1977; Lopez \& Meaburn 1986).

Nevertheless, there are a few observational clues that hint at a
possible young supernova remnant (SNR) inside Carina.  In addition to
the complex blueshifted absorption lines already noted, IR
spectroscopy with ISO revealed an unusual and strong broad 22~$\mu$m
emission feature, attributed to silicates by Chan \& Onaka (2000), who
note that the only other source where a similar emission feature is
seen is the young supernova remnant Cas~A.  Carina also displays
spatially extended diffuse soft X-ray emission across much of its
central $\sim$1$^{\circ}$ region.  The diffuse X-ray emission was
first detected in {\it Einstein} data by Seward et al.\ (1979),
although Seward \& Chlebowski (1982) concluded that the diffuse
emission could be accounted for simply by mechanical energy input from
stellar winds.  We discuss this further in Sect.\ 8.2.

\subsection{Prevailing View of Star Formation in Carina}

Early far-IR and molecular surveys described above found no
luminous, embedded sites of active star formation in the bright
central region of the nebula.  Additionally, no bright maser sources
were identified in the region (e.g., \citealt{Caswell98}).  Hence, the
prevailing view until the mid-1990s was that Carina was an evolved
H~{\sc ii} region devoid of active star formation.

This view has changed dramatically in recent years, due in part to
larger spatial coverage and higher-quality data at IR and radio
wavelengths.  In just the past decade, the Carina Nebula has been
recognized as a hotbed of active, ongoing star formation where the
star formation is occuring primarily at the periphery of the nebula
(Smith et al.\ 2000; Megeath et al.\ 1996).  It provides a laboratory
to study several star formation phenomena in great detail, all of
which have recently been identified here: (1) evaporating
protoplanetary disks (the so-called ``proplyds''), small cometary
clouds, or globules (Smith et al.\ 2003a, 2004b; Smith 2002$b$; Brooks
et al.\ 2000, 2005; Cox \& Bronfman 1995), (2) the erosion of large
dust pillars and the triggering of a second generation of star
formation within them (Smith et al.\ 2000, 2005b; Rathborne et al.\
2004; Megeath et al.\ 1996), (3) irradiated Herbig-Haro (HH) jets that
signify active accretion (Smith et al.\ 2004$a$), (4)
photodissociation regions (PDRs) on the surfaces of molecular clouds
across the region (Brooks et al.\ 2003, 2005, 1998; Rathborne et al.\
2002; Smith et al.\ 2000; Smith \& Brooks 2007; Mizutani et al.\
2004), and on the largest scales, (5) the early stages of a fledgeling
superbubble (Smith et al.\ 2000; Smith \& Brooks 2007).

Most of the current star-formation activity seems to be occurring near
the edges of the nebula, although this is not strictly true as some of
the most recent data show.  In general, the central clusters Tr~14 and
Tr~16 tend to be devoid of star formation.  DeGioia-Eastwood et al.\
(2001) and Tapia et al.\ (2003) noted the presence of young stars with
IR excess in these clusters, but these were consistent with age
spreads of several million years among the low/intermediate-mass
stellar population.  Similarly, Sanchawala et al.\ (2007) noted that
$\sim$300 X-ray sources in these clusters have properties consistent
with low/intermediate-mass pre-MS stars.  In any case, the most active
sites of ongoing star formation are not the central clusters, but the
outlying areas of the nebula.

Because of its proximity, large size on the sky, and its extreme
nature, the level of detail we can observe in the Carina Nebula makes
it appear bewilderingly complex, and it is just beginning to reveal
its secrets to us.  The following sections focus on some of the recent
work that has shed new light on Carina as an active star forming
region.

\begin{figure}[!ht]
\centering
\includegraphics[draft=False,width=0.95\textwidth]{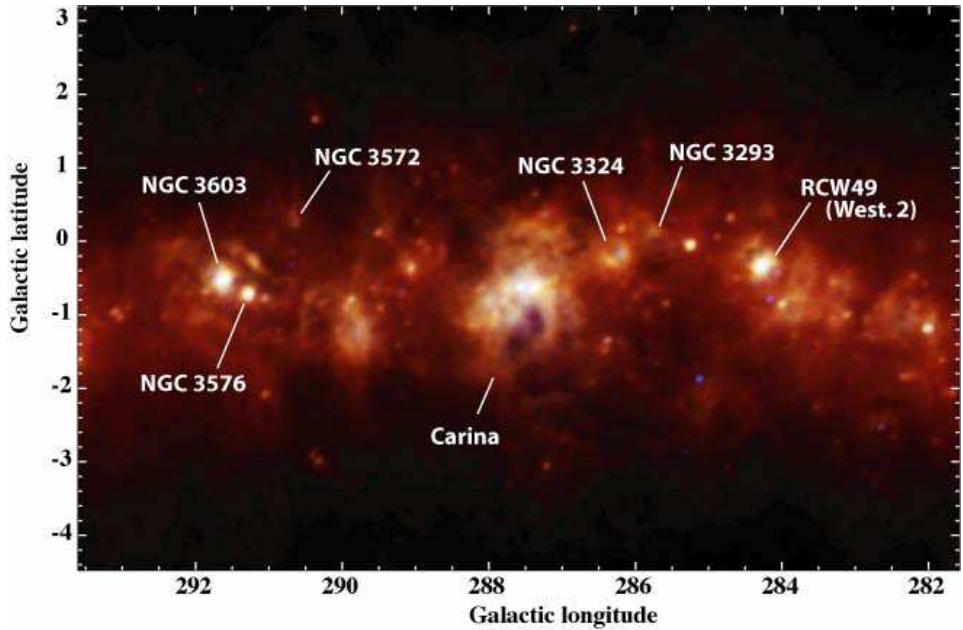}
\caption{The Milky Way in Carina observed by IRAS at 25 $\mu$m (blue),
  60~$\mu$m (green), and 100~$\mu$m (red).  Several other well-known
  regions are labeled.}
\label{fig:iras}
\end{figure}

\section{Neighborhood, Distance, and Reddening}

The Carina Nebula resides in one of the richest regions of the Milky
Way, from the point of view of massive star formation.
Figure~\ref{fig:iras} shows the part of the Galactic plane in Carina
surrounding the Carina Nebula.  This direction looks nearly down the
tangent point of the Sagittarius-Carina spiral arm, and several
massive star-forming regions are seen.  To the left of Carina (in the
direction of the Galactic center) are the rich massive cluster
NGC~3603 and the spectacular H~{\sc ii} region NGC~3576.  To the right
of Carina is RCW~49, made famous recently by Spitzer images (Whitney
et al.\ 2004) with its extremely massive cluster Westerlund 2.  Carina
is several times larger, brighter, and closer than these other
regions, many of which are at the far kinematic distance of 6--8 kpc
and highly reddened.

Unlike most massive star forming regions, the distance to the Carina
Nebula is known accurately.  A distance of 2.3 kpc (accurate to
$\pm$2\%) comes primarily from the expansion parallax of the
Homunculus nebula around $\eta$ Carinae (Smith 2006$b$, 2002$a$; Allen
\& Hillier 1993), combined with the fact that $\eta$ Car itself is
known to be within the Carina Nebula because its reflected light is
seen across the Keyhole Nebula (Walborn \& Liller 1977; Lopez \&
Meaburn 1986) and because the Homunculus occults light from the back
side of the nebula but not the front (Allen 1979).  In general, this
distance agrees with favored values derived for the star clusters in
Carina as well (Walborn 1995; Th\'{e} \& Vleeming 1971), although a
range of values up to several kpc has been reported.  Caution is
always wise, however, as the sightline toward Carina looks down a
tangent of a spiral arm as noted above.  To be more precise, this is
the distance to $\eta$~Car, the Keyhole, and Tr~16.  One normally
assumes that the other clusters described below (Tr~14, Cr~228, etc.)
are not at vastly different distances since they appear to be involved
in the same larger nebulosity and share similar kinematics.  This has
been the topic of some debate, however.

Our sightline to the Carina Nebula also suffers little extinction and
reddening compared to most massive star forming regions, affording us
the opportunity for detailed visual-wavelength studies of the stars
and fainter nebulosity.  Values of only $E(B-V)$=0.5 are favored for
the visible star clusters (Walborn 1995; Feinstein et al.\ 1973,
1980), although a casual look at an optical image reveals that
extinction is highly position-dependent in the nebula.  The reddening
law toward Carina is anomalous, with R$\simeq$4--5 instead of 3.1
(Herbst 1976; Forte 1978; Th\'{e} et al.\ 1980; Smith 1987; Smith
2002$b$).  Th\'{e} et al.\ (1980) noted that the reddening law varies
along different lines of sight in the same nebula, which continues to
dominate the uncertainty in photometric and spectroscopic distance
determinations.

\begin{figure}[!ht]
\centering
\includegraphics[draft=False,width=0.95\textwidth]{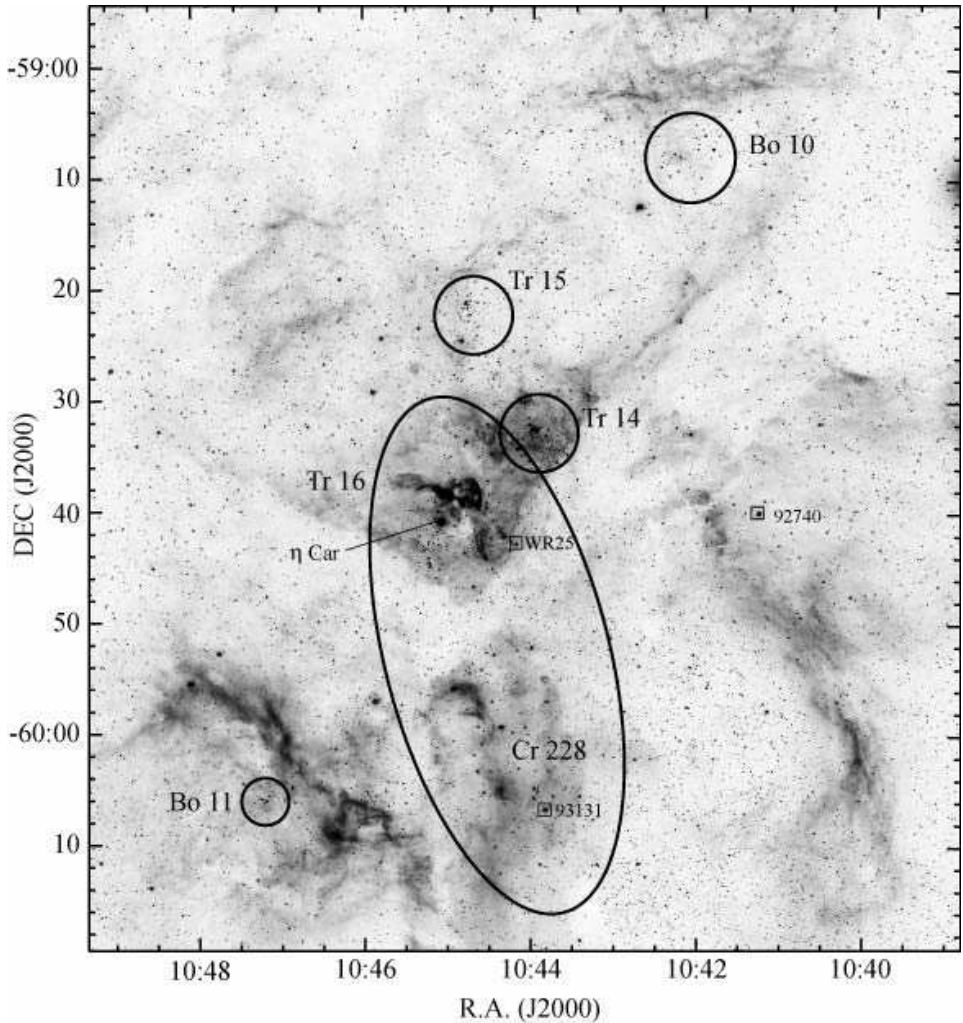}
\caption{A [S~{\sc ii}] image of the Carina Nebula identifying the
  approximate locations of the star clusters in Table 1: Trumpler 14,
  15, and 16, and Bochum 10 and 11.  Note that Collinder 228 is
  generally considered to be part of Tr~16, but these appear as two
  clusters on the sky because they are divided by an obscuring dust
  lane. The location of $\eta$ Carinae is given, and the three WNH
  stars are identified with small squares: WR25 (HD~93162), HD~93131,
  and HD~92740.  The CTIO Schmidt image is from Smith et al.\
  (2004$a$).}
\label{fig:clusters}
\end{figure}

\section{Stellar Content: Eta Carinae and its Siblings}

The Carina Nebula is probably best known as the dwelling of our
Galaxy's most luminous, massive, and unstable star -- $\eta$ Carinae.
While on the main sequence, this one star dominated the evolution of
the surrounding H~{\sc ii} region.  However, $\eta$ Car often
outshines its several spectacular neighbors that also occupy the
interior of the Carina Nebula, distributed in several sub-clusters
(Fig.~\ref{fig:clusters}).  In addition to $\eta$ Car itself, Carina
boasts an O2 supergiant (HD~93129A; Walborn et al.\ 2002$b$), three
WNH stars (late-type WN stars with hydrogen; see Smith \& Conti 2008;
Crowther et al.\ 1995) that are likely near the end of core-H burning
and are probable descendants of O2 supergiants, several O3 stars, and
more than 60 additional O stars (see Smith 2006a).  Carina also hosts
a B1.5 supergiant with a ring nebula (SBW1) that is nearly identical
to those of Sher 25 and the progenitor of SN1987A in NGC~3603 and 30
Dor, respectively (Smith, Bally, \& Walawender 2007).  This is only
the confirmed, optically-visible O star population; for example,
Sanchawala et al.\ (2007) have recently identified 16 additional OB
candidates that suffer high extinction and are only seen in the IR.
With all these O-type stars, the Lyman continuum output and total
stellar content of Carina are about the same as NGC~3603, and are
25-30\% of the Arches cluster and R136 in 30 Dor (see Smith
2006a). This fact is often overlooked in studies of massive clusters
because the O stars in Carina constitute a looser proto-OB association
(Fig.~\ref{fig:clusters}), rather than a single dense cluster.

Because of its remarkable stellar content and observability, Carina
has often been in the forefront of understanding very massive
stars. For example, Carina is the first place where the (then)
earliest spectral type O3 stars were recognized (Walborn 1973), and is
the first place where an O2~Iaf* supergiant was classified (Walborn et
al.\ 2002a).  There is much more to tell, but our goal here is to
discuss star formation in Carina, so we do not give an adequate
summary of work on the remarkable stellar content in this region; for
reviews, see Walborn (1995; 2002), Feinstein (1995), Tapia (1995), or
Massey \& Johnson (1993).

\begin{figure}[!ht]
\centering
\includegraphics[draft=False,width=0.95\textwidth]{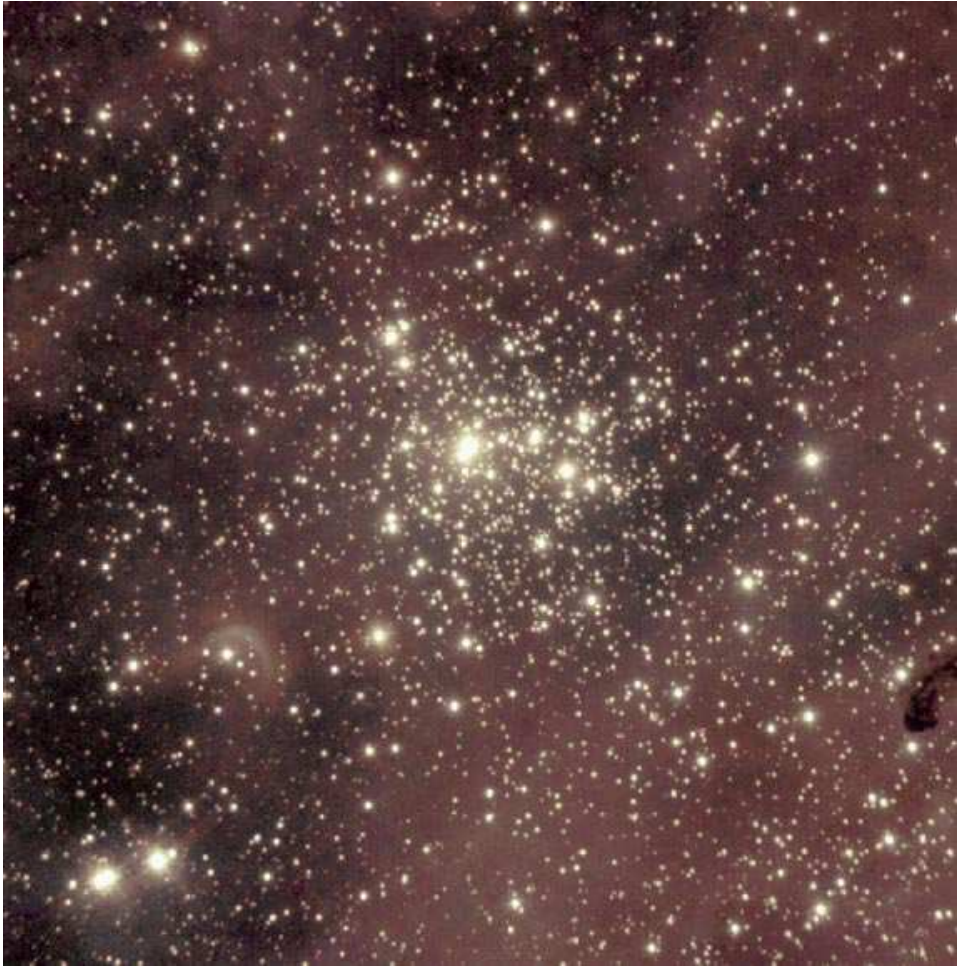}
\caption{A $JHK$ image of the Tr~14 cluster (from Ascenso et al.\
   2007).}
\label{fig:tr14}
\end{figure}

Relatively little work has been done on the low-mass pre-main-sequence
stellar population in Carina, even in the Tr~14 and 16 clusters.
Schwartz et al.\ (1990) presented a survey for H$\alpha$ emission-line
stars, but a more complete and detailed census of the young
pre-main-sequence stars is still waiting to be done.  A recent study
of deep $JHK$ photometry of Tr~14 (Ascenso et al.\ 2007) has made
significant progress in this direction (see Fig.~\ref{fig:tr14}).

Since our main focus here is on current (i.e. ongoing) star formation
in Carina, we do not review the full literature on photometry and
spectroscopy of the stars in the component clusters of the Carina
Nebula (Fig.~\ref{fig:clusters}).  Some references for each cluster
are: Tr~14 (Th\'{e} et al.\ 1980; Tapia et al.\ 1988; Morrell et al.\
1988; Massey \& Johnson 1993; Vazquez et al.\ 1996; Garcia et al.\
1998; Ascenso et al.\ 2007), Tr~15 (Th\'{e} et al.\ 1980; Morrell et
al.\ 1988; Carraro 2002), Tr~16 (Feinstein 1982; Levato et al.\ 1991;
Massey \& Johnson 1993; Kaltcheva \& Georgiev 1993), Cr~228 (Carraro
\& Patat 2001), Cr~232 (Carraro et al.\ 2004), Bo 10/11 (Feinstein
1982; Fitzgerald \& Mehta 1987).  There are also some general studies
of the stellar content of multiple clusters (Sher 1965; Moffat \& Vogt
1975; Shobbrook \& Lynga 1994; Tapia et al.\ 2004) and the larger Car
OB1 (Humphreys 1978).  Two adjacent star clusters that are not
actually part of the Carina Nebula (see Figure~\ref{fig:big}) are
NGC~3293 (Turner et al.\ 1980; Feinstein \& Marraco 1980; Shobbrook
1980) and NGC~3324 (Carraro et al.\ 2001).

We limit the remaining discussion here to the collective energy input
and feedback caused by these stars (summarized from Smith 2006a).
These stars are distributed among 5 main visually-identified clusters
listed in Table 1 and shown in Figure~\ref{fig:clusters}, although
Trumpler 16 at the center of the H~{\sc ii} region dominates all
others in terms of stellar mass and energy input.  Tr~14 is a close
second; Tr~15 and Bochum 10 and 11 are unimportant globally, but they
influence their local environments.  With these stars residing in
several loose clusters rather than one supercluster, Carina represents
the birth of an unbound OB association.  There may be a mild
age-spread in the various clusters (see Smith 2006a; DeGioia-Eastwood
et al.\ 2001).  Tr~16 is thought to be $\sim$3 Myr old, since $\eta$
Car and the three WNH stars have reached the end of their
main-sequence evolution.  Based on the magnitudes of O stars and its
compact structure, Tr~14 is somewhat younger, perhaps 1--2 Myr.  The
northern clusters Tr~15 and Bo~10 are probably older, at several Myr.
Bo~11 is very young at $\la$1 Myr, which is interesting since it is
located amid the South Pillars (see below).  Nearby, the semi-embedded
``Treasure Chest'' cluster around CPD--59$^{\circ}$2661 is probably
only $\sim$0.1 Myr old (Smith et al.\ 2005b).  Ages of these clusters,
especially Tr~14 and 16, are relevant to the properties of the ionized
and molecular gas in the region.

\begin{table}\begin{center}\begin{minipage}{5.0in}
\caption{Total Stellar Energy Input (from Smith 2006a)}
\begin{tabular}{@{}lccccc}\tableline
Cluster &Number of &log L &log $Q_H$ &$\dot{M}$ &$L_{SW}$ \\
  &O stars &(L$_{\odot}$) &(s$^{-1}$)
  &(10$^{-6}$ M$_{\odot}$ yr$^{-1}$)  &(L$_{\odot}$) \\ \tableline
Tr~16 (MS)       &47     &7.215  &50.91    &91     &45400  \\
Tr~16 (now)      &42     &7.240  &50.77    &1083   &67000  \\
Tr~14            &10     &6.61   &50.34    &18.7   &13500  \\
Tr~15            &6      &6.18   &49.56    &5.9    &1410   \\
Bo~10            &1      &6.00   &49.42    &18.3   &7120   \\
Bo~11            &5      &6.00   &49.64    &5.2    &2900   \\
(CPD-59$^{\circ}$2661)&1  &4.68   &47.88    &0.15   &33     \\ \tableline
TOTAL (MS)      &70     &7.38   &51.06    &139    &70300 \\
TOTAL (now)     &65     &7.40   &50.96    &1131   &91900 \\
%
\tableline
\end{tabular}\end{minipage}\end{center}
\end{table}

Table 1 lists the total number of O stars, bolometric luminosity,
ionizing flux, mass loss, and mechanical luminosity for each cluster,
as well as the cumulative total of all clusters (for more details see
Smith 2006a and references therein).  For the cumulative effects, two
distinct cases are important: The first case ({\bf MS}) correponds to
the history of Carina up until recent times, when $\eta$~Car was on
the main sequence, so that the massive binary was not surrounded by a
dust shell and did not have a dense LBV wind choking off its Lyman
continuum flux, and when the WNH stars in Tr~16 were presumably O2
stars as well.  For this first case, there were a total of 70 O-type
stars in Carina, producing at least $Q_H\simeq$1.2$\times$10$^{51}$
s$^{-1}$.  This would be the appropriate number to adopt when
considering the history and formation of the nebula, the lifetimes of
evaporating proplyds, globules, and dust pillars, triggered star
formation by radiative driven implosion, and the growth of the cavity
that will blow out of the Galactic plane as a bipolar superbubble
(Smith et al.\ 2000).  The second case ({\bf now}) corresponds to the
presently-observed state of the Carina Nebula, when $\eta$~Car and its
companion are surrounded by an obscuring dust shell, blocking all
their contribution to the total ionizing flux.  With $\eta$~Car and
its companion blocked by the Homunculus, the effective number of O
stars is reduced to 65 (the three WNH stars have also evolved off the
main sequence, although they still contribute very strong Lyman
continuum radiation).  For this case, the cumulative ionization source
is $Q_H$=9$\times$10$^{50}$ s$^{-1}$.  This is the number to adopt for
the {\it current} UV flux incident upon evaporating proplyds,
globules, and irradiated jets, as well as the current energy budget of
the region (Smith \& Brooks 2007).

\begin{figure}
\centering
\includegraphics[draft=false,width=\textwidth]{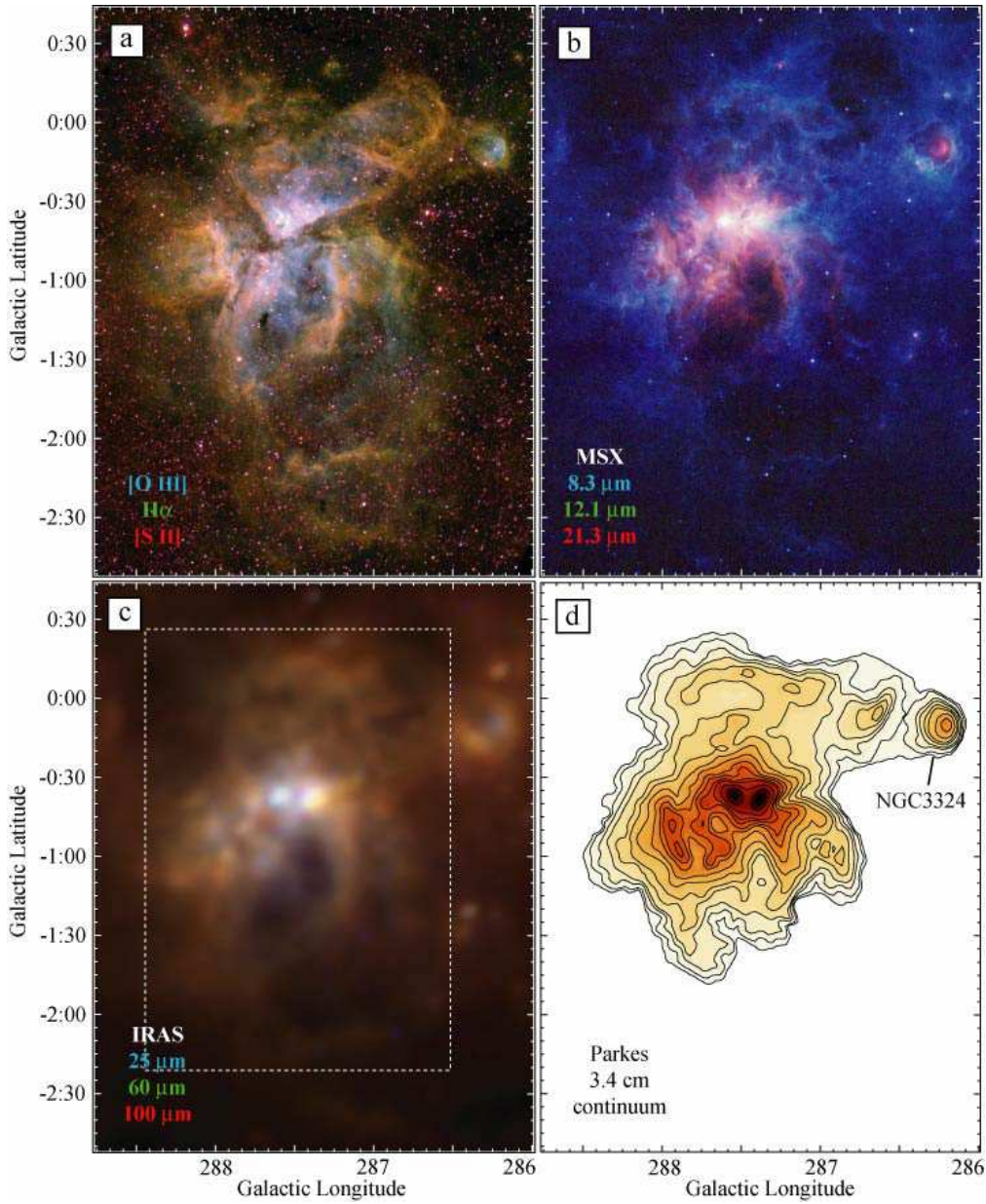}
\caption{A large-scale multiwavelength view of the entire Carina
  Nebula from Smith \& Brooks (2007). (a) Composite color visual image
  with blue = [O~{\sc iii}] $\lambda$5007, green = H$\alpha$, and red
  = [S~{\sc ii}] $\lambda\lambda$6717,6731 (from Smith et al.\ 2000).
  (b) Color image from MSX data, with blue = Band A, green = Band C,
  and red = Band E.  (c) Color image from IRAS data, with blue = 25
  $\mu$m, green = 60 $\mu$m, and red = 100 $\mu$m.  (d) False color
  image of the 3.4 cm radio continuum measured by the Parkes telescope
  (see Huchtmeier \& Day 1975).  The dashed rectangle in Panel (c) is
  the region over which Smith \& Brooks (2007) integrated the total
  flux at all wavelengths.  The axes are labeled in degrees of
  Galactic coordinates.  On the largest scales, the nebula appears to
  be bipolar, with an axis perpendicular to the Galactic plane.}
\label{fig:quad}
\end{figure}

For most of its lifetime, gas and dust in Carina has been exposed to
an ionizing luminosity about 150 times stronger than that of the Orion
Nebula.  This changed 160 years ago when $\eta$~Car ejected a thick
dust shell that cut off essentially all of its UV output, and the
total Q$_H$ of the Carina Nebula dropped by about 20\% due to the loss
of ionization from the region's most luminous member, as noted earlier
in Sect.\ 1.1.  Some dark globules seen only in silhouette today were
exposed to the harsh ionizing radiation from $\eta$ Carinae in the
past (Smith et al.\ 2003a).  This variable UV output of the central
engine makes Carina a unique laboratory for studying feedback effects.

\section{Global Properties of the Nebulosity}

Elsewhere (Smith \& Brooks 2007), we have recently summarized the
global properties of the nebula and analyzed its energy budget. The
main properties to note from that study are the following:

$\bullet$ The total IR luminosity of 1.2$\times$10$^7$ L$_{\odot}$ is
about half of the known stellar luminosity of O stars, so about half
the UV radiation from these stars escapes the giant H~{\sc ii} region.

$\bullet$ The number of ionizing photons inferred from the total radio
continuum and H$\alpha$ both underestimate the known ionizing
photon flux from the O stars (Smith 2006$a$).  Radio continuum seems to
trace $\sim$75\% of the ionizing flux, while H$\alpha$ is far worse
because of extinction, yielding a value only half that of the radio
continuum.  This might be useful for interpreting observations of
unresolved extragalactic H~{\sc ii} regions (see Smith \& Brooks 2007
for additional discussion).

$\bullet$ Emission from warm dust that dominates the SED at
$\sim$20~$\mu$m has the same spatial distribution as the radio
continuum (see Fig.\ 4$c$ of Smith \& Brooks 2007).  This warm
$\sim$80 K dust is therefore intermixed with ionized gas in the
interior of the H~{\sc ii} region, and would not be a useful tracer of
emission from star forming cores in more distant unresolved regions.

$\bullet$ Similarly, cooler dust at 30--40 K that dominates the far-IR
emission at 60--100 $\mu$m is spatially correlated with diffuse PAH
emission (see Fig.\ 4$d$ of Smith \& Brooks 2007).  We therefore
suggested that the far-IR emission traces primarily atomic gas in PDRs
at the surfaces of molecular clouds, while dust within the molecular
clouds will be even cooler, emitting primarily at submm wavelengths
where observations are still incomplete.

$\bullet$ The 30--40 K dust indicates a total emitting dust mass of
$\sim$10$^4$ M$_{\odot}$.  With a normal gas:dust mass ratio of 100,
the 10$^6$ M$_{\odot}$ of atomic gas mass located in neutral PDRs
significantly outweighs the molecular gas mass in the nebula, which is
a few times 10$^5$ M$_{\odot}$.  This is a vital clue to the current
state of star formation in an evolved H~{\sc ii} region like Carina,
because it means that the majority of the nebular mass that is present
is atomic rather than molceular, and therefore represents a huge mass
reservoir that is not being tapped for current star formation.  Smith
\& Brooks (2007) give a fuller discussion of this.

Analogous to the overall structure of 30 Doradus in the LMC, the
Carina Nebula gives us a snapshot of a giant molecular cloud being
blasted apart by UV radiation and stellar winds from the massive stars
it has given birth to.  Deep images of nebular emission at visual
wavelengths (Fig.~\ref{fig:quad}$a$; Smith et al.\ 2000) show
limb-brightened filaments and bubbles spanning more than 3$^{\circ}$
or 120 pc, with many dark obscuring dust features.  IR emission is
also seen across the same large region, mainly due to PAH emission
from PDRs at the surfaces of illuminated molecular clouds seen in the
8~$\mu$m MSX image in Figure~\ref{fig:quad}$b$, and thermal emission
from warm dust at longer wavelengths in IRAS images
(Fig.~\ref{fig:quad}$c$).  The MSX image in Figure~\ref{fig:quad}$b$,
in particular, gives the impression that the Carina Nebula consists
mainly of a large bipolar cavity that is being carved with an axis
perpendicular to the Galactic plane (Smith et al.\ 2000).  A series of
large shells seem to be breaking out of the Galactic plane toward the
lower right in this image (Smith et al.\ 2000); this is the direction
we might expect these old shells to follow if they are being distorted
by Galactic shear.

\begin{figure}[!htb]
\includegraphics[draft=False,width=\textwidth]{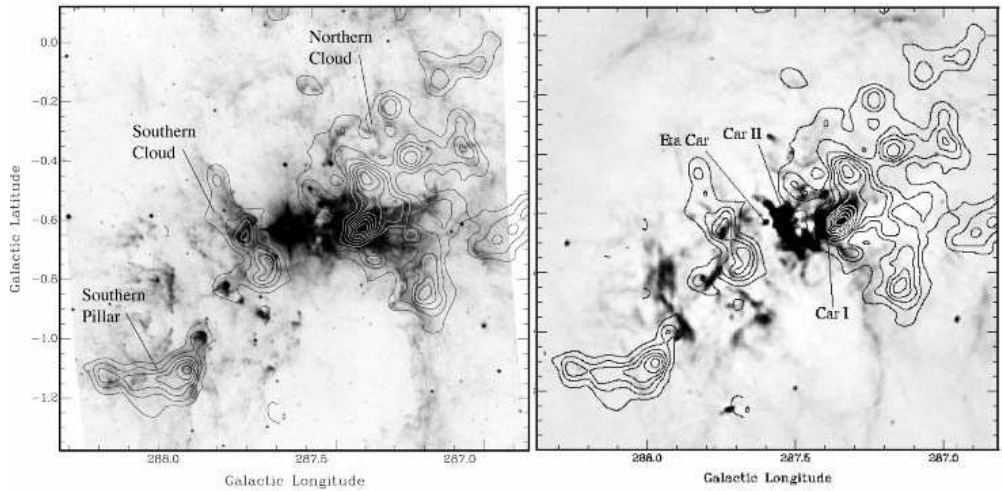}
\caption{Molecular gas distribution in the Carina Nebula, with
  contours from the NANTEN $^{12}$CO(1--0) survey of Yonekura et al.\
  (2005).  These contours are superposed on images at 8~$\mu$m from
  MSX (left; Smith et al.\ 2000) and in the 0.843 GHz continuum from
  MOST (right; Whiteoak 1994).}
\label{fig:co}
\end{figure}


The pair of images in Figure~\ref{fig:co} demonstrate the relationship
between the molecular cloud, the ionized gas and the IR dust emission
across the brightest part of the nebula.  The left panel in
Figure~\ref{fig:co} shows a zoomed-in view of the 8~$\mu$m MSX
emission described earlier. The image is dominated by the spectacular
southern dust pillars, some with bright compact emission sources at
their tips. The two radio continuum sources Car I and Car II can be
identified in the right panel of Figure~\ref{fig:co}. This 0.843 GHz
(36~cm) radio continuum image \citep{Whiteoak94} has been taken from
the Molonglo Galactic Plane Survey (MGPS) made with the Molonglo
Observatory Synthesis Telescope (MOST). The contours overlayed on both
the MSX and MOST images represent the NANTEN $^{12}$CO data at 115 GHz
with a 2.7\arcmin\ beam \citep{Yonekura05}. The data have been
integrated over the velocity of range --25 to --15 \kms (LSR), where
the bulk of the molecular gas is found (Lee et al.\ 2000). The
molecular gas is concentrated into three components that have been
labeled ``Northern Cloud'', ``Southern Cloud'' and ``Southern Dust
Pillar''. The first two components are the same northern and southern
cloud components identified in the previous CO survey work of
\citet{deGraauw81}, \citet{Whiteoak84} and more recently
\citet{Brooks98}. The third molecular component traces the giant
pillar first identified by \citet{Smith00}.

Figure~\ref{fig:co} illustrates the close correspondence between the
molecular gas and the MSX 8~$\mu$m emission that exists throughout the
nebula, and in particular at the edges of the molecular clouds facing
$\eta$~Car.  This supports the finding by \citet{Zhang01} for
wide-spread PDR emission throughout Carina. Using the Antarctic
Submillimeter Telescope and Remote Observatory (AST/RO),
\citet{Zhang01} obtained fully sampled maps of the Carina GMC complex
in the CO(4--3) and [C~{\sc i}] transitions at 460 GHz and 492 GHz,
respectively, with a $\sim$3\arcmin\ (FWHM) beam. Both of these
transitions probe warm (T$\simeq$50~K) and dense ($n_e > 10^3$
cm$^{-3}$) gas associated with PDRs. Using the CO(1--0) and CO(4--3)
data they derived average excitation temperatures of 30--50 K.
Further confirmation of the widespread PDR emission in Carina comes
from the results of 3.29~$\mu$m PAH observations using the SPIREX/Abu
telescope at the South Pole \citep{Rathborne02}. A well established
tracer of PDR emission, the 3.29~$\mu$m PAH emission correlates very
well with MSX 8~$\mu$m PAH emission.  As noted earlier, Smith \&
Brooks (2007) showed that the PAH emission is, in turn, well
correlated with the 30--40~K dust emission that dominates the far-IR
SED, and the neutral gas mass of 10$^6$ traced by that dust represents
the dominant phase of nebular material in Carina.

From the point of view of star formation occurring in molecular
clouds, this global view of Carina suggests three main areas of
interest, sampling three different phases in the development of a GMC:
1) The Keyhole region and its small molecular globules represent the
last shredded bits of a GMC core that has already formed a massive
cluster; 2) the northern cloud near the Tr~14 cluster represents a
relatively pristine GMC that does not yet seem to have very active
star formation, but is clearly being irradiated by a massive cluster;
and 3) the southern cloud and south pillars represent a region that is
currently being eroded and shaped by radiation and winds from a young
massive cluster, giving rise to numerous dust pillars with a second
generation of stars forming within them.  These three regions will be
discussed in the following sections.

\begin{figure}[!ht]
\includegraphics[draft=False,width=\textwidth]{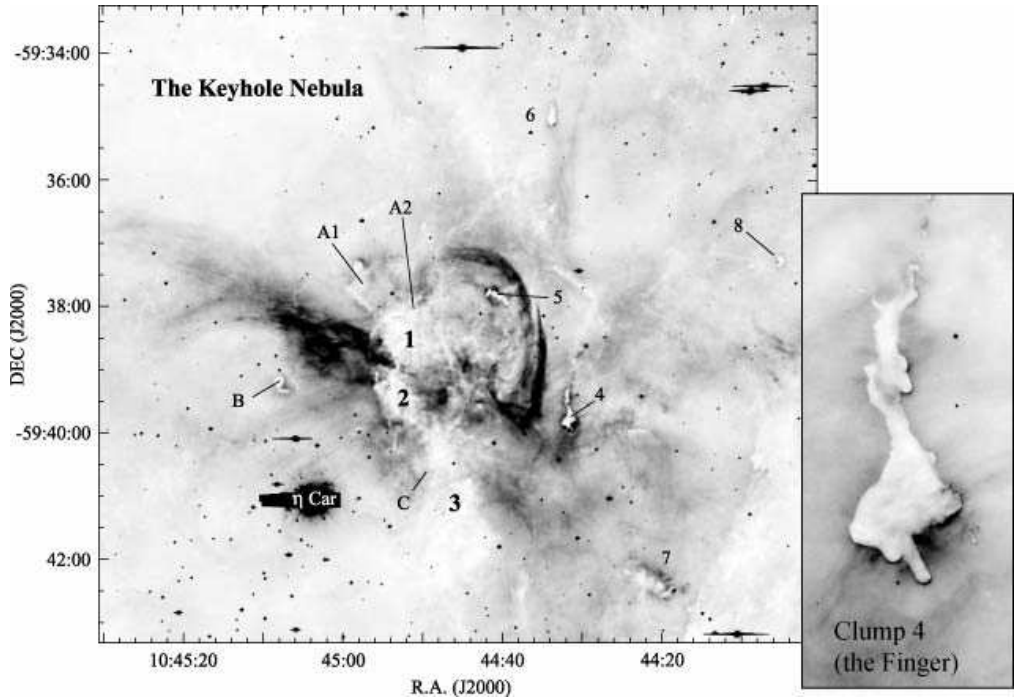}
\caption{An H$\alpha$ image of the Keyhole nebula and its molecular
  globules, obtained with the MOSAIC2 camera on the CTIO 4m telescope
  (see Smith et al.\ 2003a); see Table 2 for details.  The detail of
  the ``Finger'' (clump 4) is an H$\alpha$ image obtained with
  HST/WFPC2 (Smith et al.\ 2004b).}
\label{fig:keyhole}
\end{figure}

\section{Main Regions of Interest I:  The Keyhole and Tr~16}

The Keyhole Nebula, located in the center of the larger Carina Nebula
about 3\arcmin\ from $\eta$ Carinae within the Tr~16 cluster, is one
of the brightest emission-line nebulae in the sky
(Fig.~\ref{fig:keyhole}).  In fact, the ionized gas emission is so
bright that the strong 1.2~mm continuum emission is dominated by
free-free, not cool dust (Brooks et al.\ 2005) -- a property that is
rare among star-forming regions. It has a strange arc-shaped structure
superposed with many dark obscuring clumps that give it the appearance
of an old-fashioned keyhole shape.  The velocity field in H110$\alpha$
emission does not match the simple picture of an expanding bubble
centered on $\eta$ Car (Brooks et al.\ 2001), as had been suggested
from earlier work (see Sect.\ 1.2).  Despite its prominence, the
physical origin of this structure remains a mystery, although a direct
association with $\eta$ Car has been proposed (Smith 2002$b$).  Many
small molecular globules scattered around the Keyhole (Cox \& Bronfman
1995; Brooks et al.\ 2000) suggest that the Keyhole and its associated
globules may be the last vestiges of the original molecular cloud core
that spawned Tr~16.

\begin{table}\begin{center}\begin{minipage}{3.7in}\footnotesize
\caption{Molecular globules around the Keyhole}
\begin{tabular}{@{}lcccc}\tableline
Clump &$\alpha_{2000}$ &$\delta_{2000}$ &M/M$_{\odot}$ &Comment  \\ \tableline
A1  &10 44 57.5  &--59 37 40  &14-19 &Kangaroo Nebula  \\
A2  &10 44 51.5  &--59 38 00  &16    &...  \\
B   &10 45 07.9  &--59 39 13  &3-5   &...  \\
C   &10 44 50.4  &--59 40 35  &6-8   &...  \\
1   &10 44 52.8  &--59 38 27  &17    &Dark cloud  \\
2   &10 44 53.4  &--59 39 25  &11    &Dark cloud  \\
3   &10 44 46.0  &--59 41 30  &4     &Dark cloud  \\
4   &10 44 31.4  &--59 38 48  &5-6   &The Finger  \\
5   &10 44 40.4  &--59 37 51  &1     &...  \\
6   &10 44 33.8  &--59 35 00  &5-10  &...  \\
7   &10 44 20.5  &--59 42 26  &3-12  &...  \\
8   &10 44 05.1  &--59 37 16  &9-91  &...  \\

\tableline
\end{tabular}\end{minipage}\end{center}
\end{table}

These small Bok globules are analogous to Thackeray's globules in
IC~2944 (Thackeray 1950a; Reipurth et al.\ 2003).  They have typical
sizes of 0.1--1 pc and masses of a few to 10 M$_{\odot}$, as listed in
Table 2 (see Cox \& Bronfman 1995; Brooks et al.\ 2000, 2005; Smith
et al.\ 2004$b$; Smith 2002$b$).  Whether or not they are sites of
ongoing or potential future star formation remains an open question.
No embedded IR sources inside these Keyhole globules have been
reported yet, but a detailed analysis of one source (the so-called
``Finger''; see Fig.~\ref{fig:keyhole}) indicates that it is currently
experiencing external overpressure and may therefore be encountering
radiative-driven implosion (Smith et al.\ 2004$b$).  The globules are
massive enough and in a harsh-enough environment that they are good
candidates for sites of future triggered star formation.

The enigma of the dark keyhole structure dates back to 1830s and the
first detailed drawings of the Keyhole by Herschel (see Sect.\ 1.1 and
Fig.~\ref{fig:herschel}). \citet{Cox951} detected molecular emission
from three dark clumps labeled clumps 1, 2 and 3 in
Figure~\ref{fig:keyhole}, with mass estimates from \CO\ observations
of 17, 11, and 4 \Msun, respectively. Unlike the other molecular
condensations, clumps 1, 2, and 3 lack bright H$\alpha$ emission rims
and PDR emission (\citealt{Brooks00}; \citealt{Rathborne02}; Smith
2002b; Walborn 1975).  Nor were they detected in the 1.2-mm continuum
(Brooks et al.\ 2005), suggesting lower temperatures than the other
clumps. For a clump of 10 \Msun\ and $T_d=10$ K, we expect a flux
density at 1.2 mm of 0.033 Jy, equivalent to the $2\sigma$ observed
detection limit. The absence of 1.2-mm emission requires that the dark
Keyhole structure is comprised of cool ($T_d < 15 $ K) foreground
molecular clumps at the outskirts of the \HII\ region
\citep{Brooks00}.  The three dark Keyhole clumps may provide a
snapshot of an earlier phase that the bright-rimmed molecular globules
in the nebula have already passed through -- before they were pressure
confined and externally photoevaporated. This is supported by the fact
that the dark Keyhole clumps are bigger, more massive, darker, and
more diffuse than the others. The bright-rimmed globules are being
photoevaporated by radiation from Tr~16, while the dark clumps are
somehow shielded even though they are projected in almost the same
place.

By contrast, Carina also contains many smaller condensations and
cometary clouds -- the so-called proplyd candidates (Smith et al.\
2003a).  These much smaller objects are probably in a {\it more}
advanced stage of photoevaporation, with only their dense
circumstellar envelopes remaining.  It will be an interesting goal for
future studies to determine if these objects harbor young stars.  The
proplyd candidates are scattered throughout the H~{\sc ii} region, and
some are bright objects while others are seen only in silhouette.

\begin{figure}[!ht]
\includegraphics[draft=False,width=\textwidth]{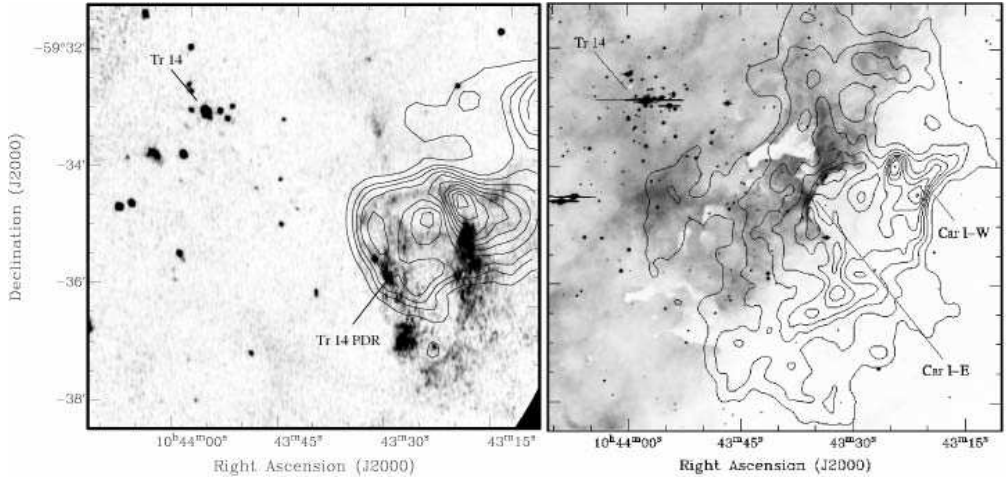}
\caption{The edge-on PDR near Tr~14.  {\it Left}: $^{13}$CO(2--1)
  contours (Brooks et al.\ 2003) on 3.3~$\mu$m PAH emission (Rathborne
  et al.\ 2002).  {\it Right}: 4.8 GHz radio continuum contours
  (Brooks et al.\ 2001) on an H$\alpha$ image.}
\label{fig:pdr}
\end{figure}

\section{Main Regions of Interest II:  The Edge-On PDR Near Tr~14}

From the distribution of molecular gas shown in Figure~\ref{fig:co} it
is evident that the molecular cloud in the vicinity of Tr~14 is more
intact than the remaining fragments in the Keyhole. In this so-called
northern molecular cloud, the brightest molecular emission is
concentrated toward the dark western dust lane seen in visual images,
offset from the center of Tr~14 by $\sim$4\arcmin. The radio continuum
emission source Car I is also located here, centered on the interface
of the dust lane and the bright H~{\sc ii} region. Between Car I and
the molecular cloud, widespread PDR emission is seen in arc-like PAH
emission features at 3.3~$\mu$m \citep{Rathborne02}.  The spatial
sequence of Tr~14, Car I, PAH emission, and then strong molecular-line
emission delineates a classical edge-on PDR \citep{Brooks03}.

Figure~\ref{fig:pdr} shows the spatial relationship of Tr~14, the
ionized gas, the PDR emission, the molecular gas, and the dark dust
lane.  The curved edges of both $^{13}$CO and 3.3~$\mu$m PAH emission
correspond to a bright optical rim at the edge of the dark lane in
H$\alpha$, and match a curved radio continuum arc identified as
Car~I-E (see Fig.~\ref{fig:pdr}).  There is a clear detection of the
CS(5--4) line here, which indicates the presence of high density gas
(\citealt{Brooks03}), near the critical H$_2$ density of
n$_{cr}$(H$_2$)=8$\times$10$^6$ cm$^{-3}$ for this transition.
Curving in the same direction but located at the western edge of the
continuum emission is a second arc, labeled Car I-W. Its sharp edge
marks a western boundary of the extended radio continuum.  Two compact
H~{\sc ii} regions have also been identified with fluxes that
correspond to ionization by single O9.5 and B0-type stars, which may
trace recent star-forming activity in the northern cloud.  Tapia et
al.\ (2006, 2003) have also identified potential sites of ongoing star
formation here.

The northern molecular cloud appears to wrap around Car I, suggesting
that Tr~14 may be carving out a small ionized cavity. H110$\alpha$
recombination-line data do not exhibit the characteristic
double-peaked profiles of expansion, but if the expansion velocity is
smaller than the sound speed of the ionized gas, the profiles of Car I
would be consistent with an age $<$10$^6$ yr (\citealt{Brooks03}).

To study the properties of the PDR at the interface of the dust lane,
\citet{Brooks03} utilized observations of the fine-structure emission
lines [\CII] 158 $\mu$m and \OI \ 63 $\mu$m from the Kuiper Airborne
Observatory (KAO), as well as $43-196$ $\mu$m data from the Long
Wavelength Spectrometer (LWS) onboard ISO. Strong [\CII] 158 $\mu$m
and [\OI] 63 $\mu$m emission is concentrated toward the dark dust
lane, in good correspondence with the molecular gas. Results from a
1-dimensional PDR model toward the [\CII] 158 $\mu$m emission peak
imply a density of $10^4$ cm$^{-3}$ and a FUV field of $10^4$ G$_0$,
where G$_0$=1.6$\times$10$^{-3}$ ergs s$^{-1}$ cm$^{-2}$ is the Habing
flux (Habing 1968), representing the average diffuse FUV flux of the
local interstellar radiation field. Using the ISO spectrum, estimates
of the dust temperatures were found to be $40-50$ K towards the PDR
and the molecular cloud and 65 K toward the \HII\ region. Such a
temperature gradient was first noted by \citet{Cox952} using
high-resolution images of the Carina Nebula in the four IRAS bands.
The clear tendency for the IR distribution to shift away from Tr~14 as
one goes from the mid-IR to far-IR signifies a strong temperature
gradient, likely caused by external heating from the massive stars in
Tr~14.

At a projected distance of $\sim$2 pc, the UV output of Tr~14 dominates
over the other Carina Nebula clusters like Tr~16 in determining the
local FUV flux at the PDR in the northern cloud (\citealt{Brooks03};
\citealt{Smith06}).  The emanating radiation field is sufficient to
produce the ionization fronts and the FUV flux measured at the PDR
emission peak. Furthermore, the kinetic energy provided by the stellar
winds can sustain the high velocity dispersion measured in the
molecular gas \citep{Brooks03}.  A value of $\sim$10$^4$ G$_0$ from
the Tr~14 FUV field is comparable to that found in 30 Dor, but weaker
than the $\sim$10$^5$ G$_0$ fields in regions such as M17, the Orion
Bar, and W49N. In Orion, the characteristics of the O6-O7 star
$\theta^1$Ori~C determine most of the properties of the ionized
material and the PDR (\citealt{ODell01}; \citealt{Hollenbach97}). The
distance between $\theta^1$Ori~C and the main ionization front is
$\sim$0.25 pc, while the distance between Tr~14 and Car I is $\sim$2
pc. Moreover, the ionization front adjacent to the brightest [\CII]
emission in 30 Dor is located $\sim$20 pc from the luminous cluster
R136 \citep{Israel96}. Thus, the greater displacement between the
ionization fronts and the exciting clusters in Tr~14 and 30 Dor
explains why the measured PDR FUV fields are lower.  Evidently, Tr~14
has been more effective or has had more time to clear away its
immediate birth environment (Tr~14 is probably 1--2 Myr old, while
$\theta^1$Ori~C in the Trapezium is only of order 0.1--1 Myr old).

\begin{figure}[!ht]
\includegraphics[draft=False,width=\textwidth]{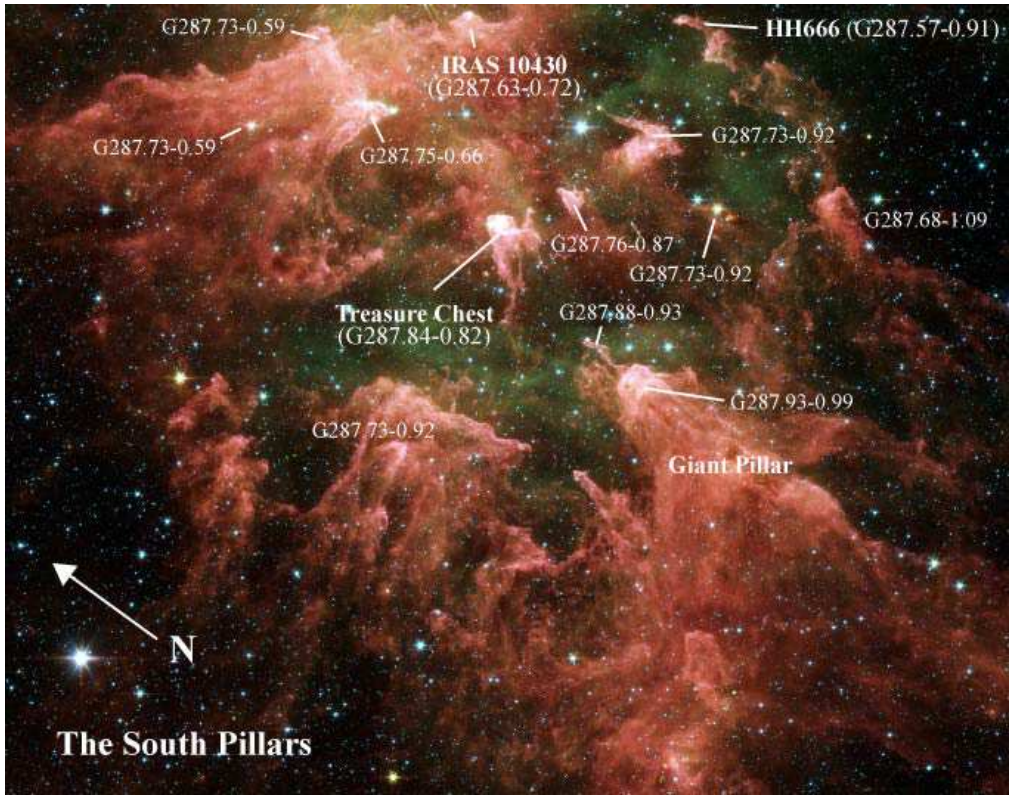}
\caption{Spitzer/IRAC image of the South Pillars (Smith et al., in
  prep).}
\label{fig:spitzer1}
\end{figure}

\section{Main Regions of Interest III:  Star Formation in the South Pillars}

In the past few years, it has become clear that the most active region
of ongoing star formation in the Carina Nebula is the so-called
``South Pillars'', where winds and UV radiation from Tr~14 and 16 are
sweeping through and destroying a GMC.  From a census of the stellar
energy budget in the region (Smith 2006a), the FUV field in the south
pillars is milder than in the central parts of the nebula; it falls in
the range of roughly 500--5000 G$_0$.  The significance of this region
was first recognized after wide field thermal-IR images from the {\it
MSX} satellite became available (Smith et all.\ 2000; see
Fig.~\ref{fig:quad}$b$), changing the predominant view that Carina was
an evolved H~{\sc ii} region without active star formation.  Since
then, it has become the main focus for studies investigating ongoing
and triggered star formation in Carina.  A recent Spitzer/IRAC image
of the South Pillars is shown in Figure~\ref{fig:spitzer1}.  It is
clear from this image that there are many interesting sites of ongoing
star formation in addition to the few case studies described below,
which are so far the ones that have been examined in detail.

\begin{table}[p]
\begin{center}\begin{minipage}{4.25in}
\caption{Embedded star formation sites in Carina}
\begin{tabular}{@{}lccc}\tableline
Source &$\alpha_{2000}$ &$\delta_{2000}$ &Other name  \\ \tableline
G287.22-0.53  &10 42 49.2  &--59 25 27  &...  \\
G287.57-0.91  &10 43 51.3  &--59 55 22  &HH 666  \\
G287.68-1.09  &10 43 57.2  &--60 08 25  &...  \\
G287.87-1.36  &10 44 17.9  &--60 27 46  &IRAS 10423-6011  \\
G287.47-0.54  &10 44 32.8  &--59 33 20  &N4  \\
G287.73-1.01  &10 44 34.6  &--60 05 36  &HD~93222  \\
G287.73-0.92  &10 44 45.4  &--59 59 16  &...  \\
G287.51-0.49  &10 44 59.0  &--59 31 24  &...  \\
G287.63-0.72  &10 45 01.1  &--59 47 06  &IRAS 10430-5931  \\
G287.52-0.41  &10 45 20.9  &--59 27 28  &IRAS 10434-5911  \\
G287.76-0.87  &10 45 22.3  &--59 58 23  &...  \\
G287.84-0.82  &10 45 53.6  &--59 57 10  &Treasure Chest  \\
G287.93-0.99  &10 45 56.1  &--60 08 50  &Giant Pillar  \\
G287.88-0.93  &10 46 00.9  &--60 05 12  &IRAS 10441-5949  \\ 
G287.75-0.66  &10 46 01.2  &--59 46 58  &...  \\
G287.73-0.59  &10 46 08.0  &--59 42 41  &...  \\
G287.86-0.82  &10 46 13.9  &--59 58 41  &...  \\
G287.80-0.56  &10 46 46.8  &--59 43 25  &...  \\
G288.07-0.80  &10 47 35.4  &--60 02 51  &...  \\
\tableline
\end{tabular}\end{minipage}\end{center}
\end{table}

\smallskip
\noindent $\bullet$\ {\it IRAS 10430-5931:} A study of this embedded
IR source by Megeath et al.\ (1996) provided the first strong evidence
that active star formation was still happening in the Carina Nebula.
This source is embedded in a bright-rimmed globule at the inner edge
of the dark obscuring dust lane that bisects the Carina Nebula, and
points toward $\eta$ Car (see Fig.~\ref{fig:spitzer1}).  Megeath et
al.\ estimated that this globule has a luminosity of roughly 10$^4$
L$_{\odot}$ with $\sim$70 M$_{\odot}$ of gas, possibly containing
several embedded point sources.  The recognition that this was a site
of active star formation at the edge of the molecular cloud south of
$\eta$ Car suggested that astronomers needed to be looking farther
away from the core of the nebula to the south and southeast in order
to identify additional sites of star formation.  This work has just
begun, and a few such sources are identified below.

\begin{figure}[!p]
\includegraphics[draft=False,width=\textwidth]{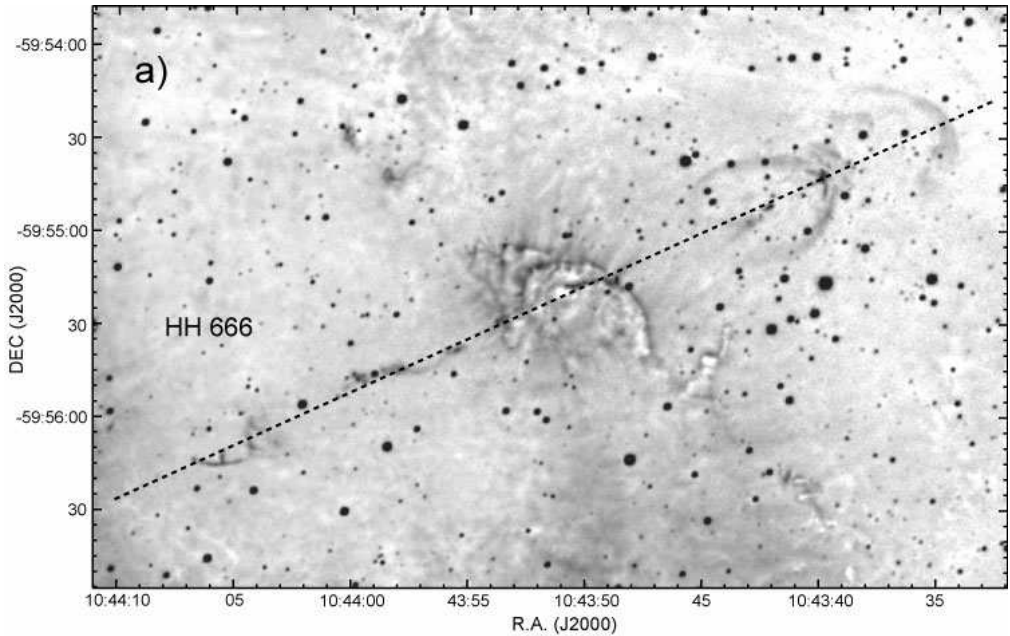}
\caption{The Axis of Evil in the Carina Nebula.  HH~666 is the first
  Herbig-Haro jet to be identified here (from Smith et al.\ 2004$a$).}
\label{fig:hh666}
\end{figure}

\smallskip
\noindent $\bullet$\ {\it HH~666 - The Axis of Evil:} One definite
signpost of active star formation is the presence of protostellar
outflows, or Herbig-Haro jets, that punch out of a molecular cloud
core where young embedded stars are actively accreting material from
their circumstellar disk.  The first protostellar outflow to be
identified in Carina is HH~666, dubbed the ``Axis of Evil'' (Smith et
al.\ 2004$a$).  This is a remarkably straight, several parsec-long
bipolar jet (see Fig.~\ref{fig:hh666}) emanating from a molecular
globule G287.57-0.91 in the South Pillar region.  In the near-IR, the
jet exhibits extremely bright [Fe~{\sc ii}] emission, but is not seen
in H$_2$.  The embedded source is likely to be an
intermediate-luminosity ($\sim$200 L$_{\odot}$) Class I protostar
called HH~666~IRS, located at 10$^h$43$^m$ 51$\fs$3,
--59$^{\circ}$55\arcmin21\arcsec\ (J2000).  This jet was identified
on ground-based images, and was seen because it is the most
spectacular jet in the region, and one of the longest HH jets known to
date.  It is likely that many other HH jets are still waiting to be
discovered in Carina (see preliminary results from HST below).

\begin{figure}[!ht]
\centering
\includegraphics[draft=False,width=0.9\textwidth]{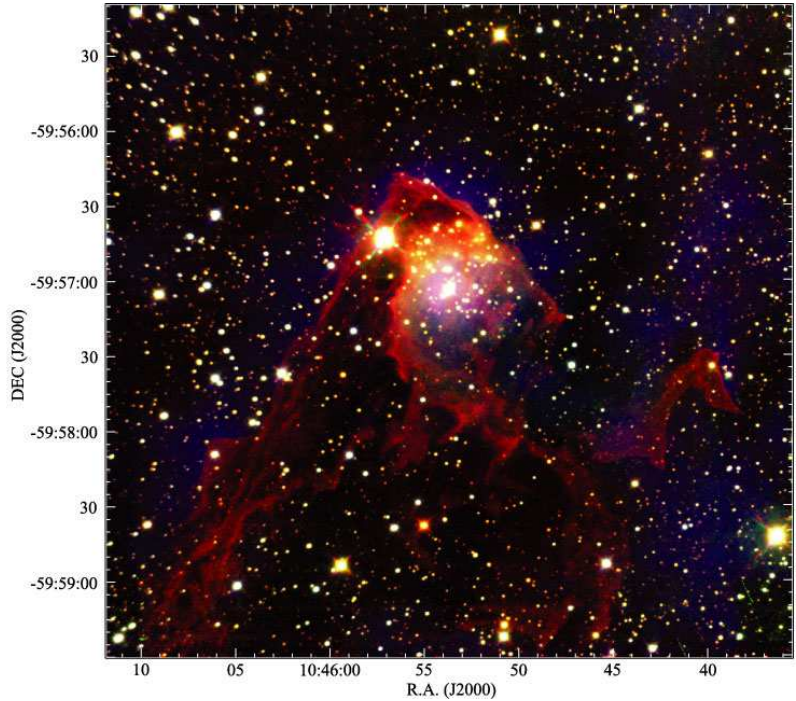}
\caption{Narrowband near-IR image of the Treasure Chest, an embedded
  cluster in the South Pillar region.  Blue is Pa$\beta$, green is
  [Fe~{\sc ii}] $\lambda$16435, and red is H$_2$ (see Smith et al.\
  2005).}
\label{fig:treasure}
\end{figure}

\smallskip
\noindent $\bullet$\ {\it The Treasure Chest:} In addition to
individual young stars embedded within the heads of dust pillars, the
South Pillar region also contains young embedded clusters.  The
brightest and most spectacular of these (and the only one yet to be
studied in detail) is the Treasure Chest cluster associated with the
dust pillar G287.84-0.82 (Smith et al.\ 2005; H\"{a}gele et al.\ 2004;
Rathborne et al.\ 2004).  A narrowband near-IR image of the Treasure
Chest is shown in Figure~\ref{fig:treasure}, where the young star
cluster is seen to be carving out a cavity inside the head of a dust
pillar, the outside of which has a bright and spectacular PDR seen in
H$_2$ emission.  The most luminous member of ths star cluster is the
O9.5 V star CPD$-$59$^{\circ}$2661. It can be seen at optical
wavelengths, surrounded by a small fuzzy H~{\sc ii} region that has
been known for some time (Thackeray 1950b; Walsh 1984).  Smith et al.\
(2005b) found that this embedded cluster (with $A_V$ as high as 50)
was very young, with a likely age less than 0.1 Myr.  They also found
that it has one of the highest disk fractions of any known embedded
cluster; the disk fraction is 67\% measured from near-IR excess in the
$K$ band.  Higher disk fractions generally result when
longer-wavelength data are included.  For example, the young cluster
NGC~2024 has a disk fraction of 60\% measured in JHK photometry (lower
than Carina's Treasure Chest), but a higher disk fraction of 90\% when
L-band photometry is included (Haisch et al.\ 2000).  Thus, the true
disk fraction in Carina's Treasure Chest may be near 100\%, making it
a valuable target for learning about the effect of UV irradation from
nearby O stars on young protoplanetary disks.

\smallskip
\noindent $\bullet$\ {\it The Giant Pillar:} Among the South Pillars,
the largest one is a giant dust pillar more than 25 pc across that
points toward $\eta$ Carinae (Smith et al.\ 2000).  Many smaller dust
pillars, more typical of those seen in other H~{\sc ii} regions, are
seen to sprout from this one giant pillar, which also contains several
embedded IR sources (Fig.~\ref{fig:spitzer1}; Rathborne et al.\ 2004).
This giant pillar is probably the remains of a GMC core that is being
shredded by feedback from the stars in Tr~16.  This and several
additional pillars in this region await detailed study.

In addition to these, there are many other sites of ongoing star
formation in the South Pillars region and around the Carina Nebula
(see Figures~\ref{fig:quad}$a$ and \ref{fig:spitzer1}).  A list of
several mid-IR sources with embedded stars is given in Table 3,
compiled from Rathborne et al.\ (2004) and Smith et al.\ (2000).  Some
of these are potentially sites of embedded, ongoing {\it massive} star
formation.

\begin{figure}[!p]
\includegraphics[width=0.98\textwidth,draft=False]{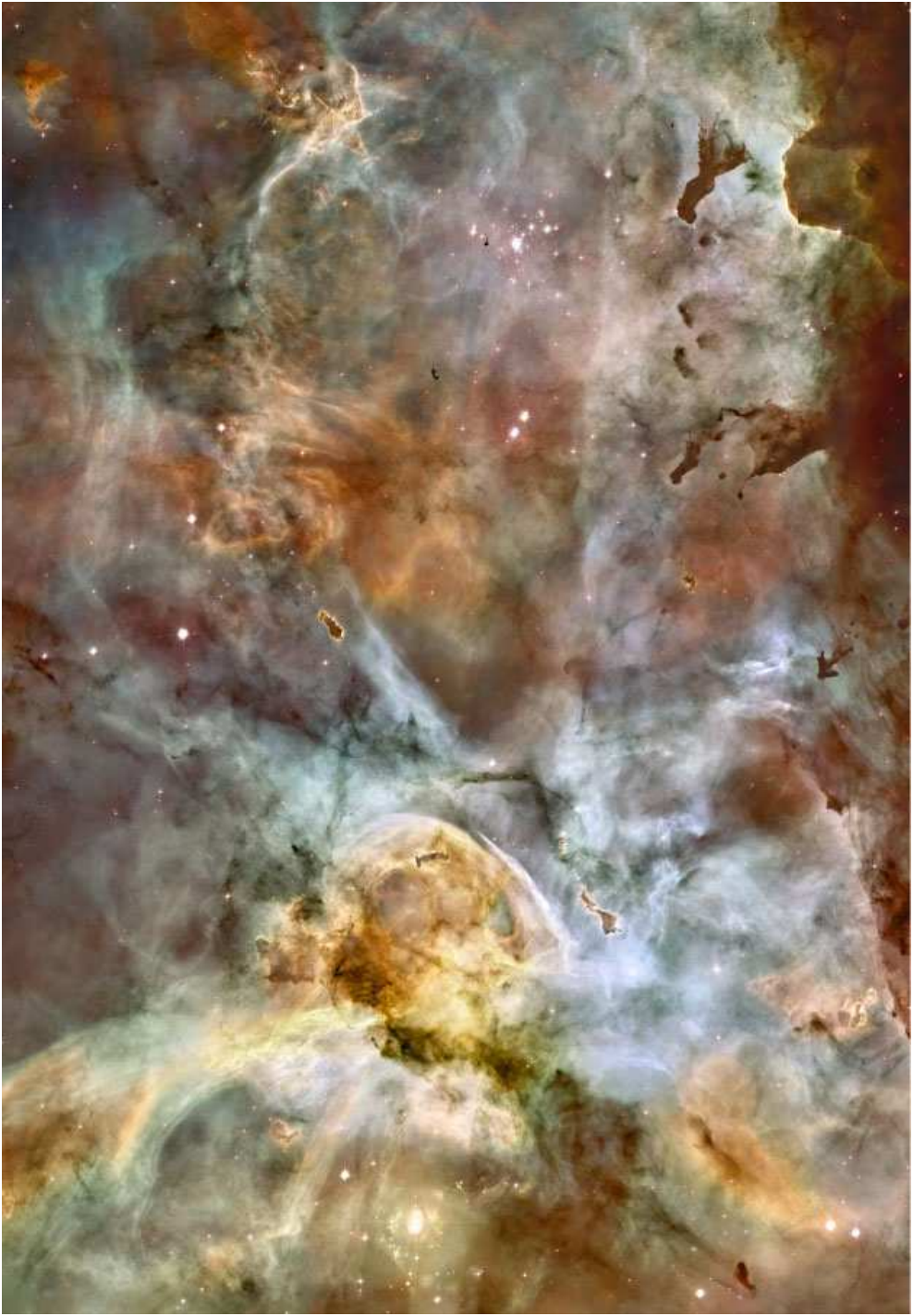}
\caption{The recent HST/ACS-H$\alpha$ imaging survey of the Carina
  Nebula (PI: Smith), from a mosaic released by the Hubble Heritage
  Team.  The color map is taken from ground based images in the usual
  filters with [O~{\sc iii}]=blue, H$\alpha$=green, and [S~{\sc
  ii}]=red.  $\eta$~Car is at the very bottom. {\it Image credit}:
  N. Smith, NASA, ESA.}
\label{fig:heritage}
\end{figure}
\begin{figure}[!ht]
\centering
\includegraphics[width=\textwidth,draft=False]{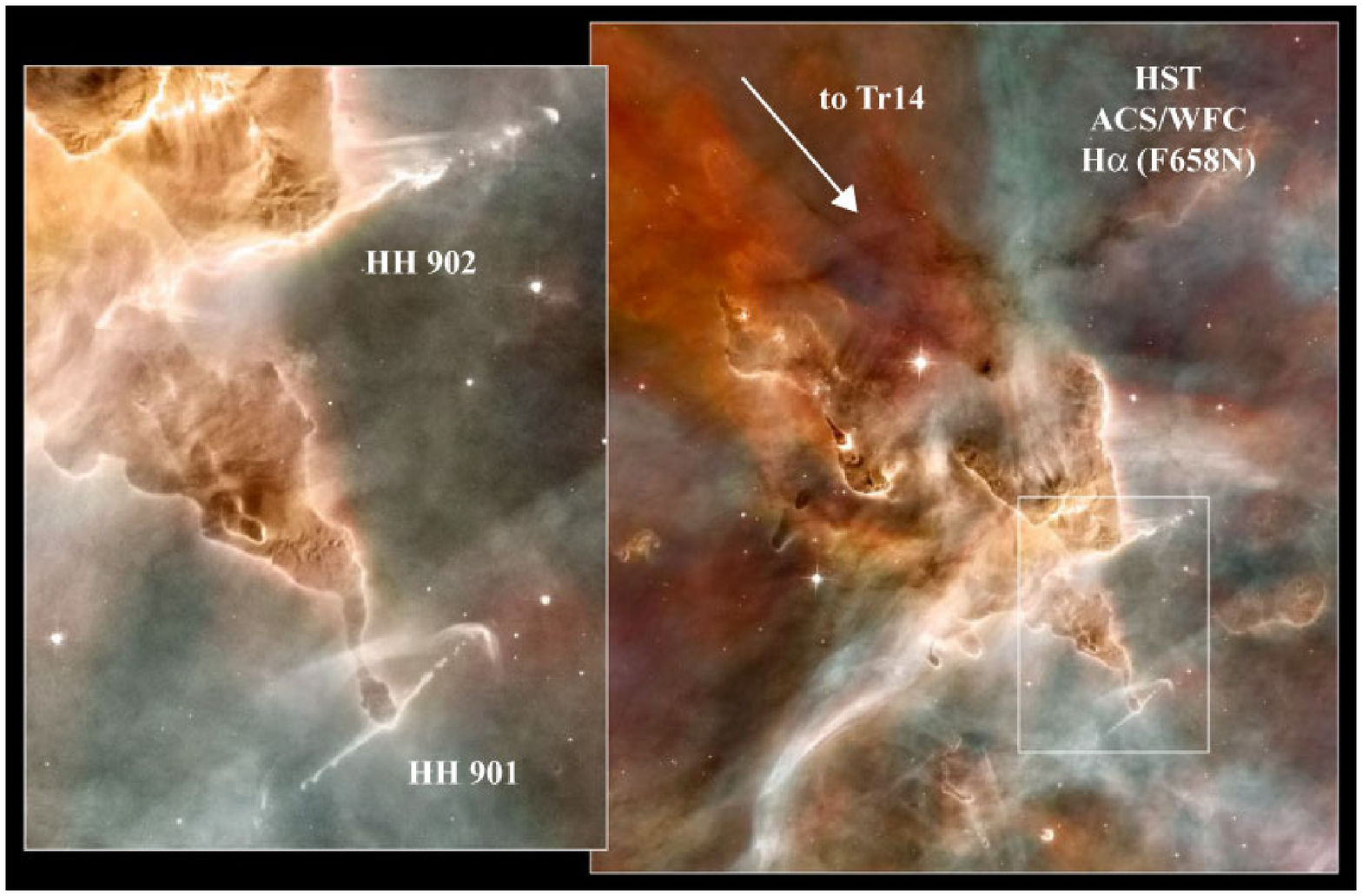}
\caption{One small section of the HST images in the previous figure.
  This detail shows a region about 10\arcmin\ north of the Tr~14
  cluster, containing a dust pillar in the process of being shredded.
  Two newly discovered bipolar Herbig-Haro jets, HH~901 and 902,
  emerge from Class~I protostars buried within the heads of the dust
  pillars (Smith et al.\ in prep.).}
\label{fig:jets}
\end{figure}

\section{A Preview of Coming Attractions}

\subsection{An {\it HST}/ACS H$\alpha$ Survey of Carina}

At the time of writing this review, the first large imaging survey of
the Carina Nebula with the Hubble Space Telescope (HST) has just been
completed, and so far the results are stunning.  The only previous HST
imaging of Carina involved smaller WFPC2 fields of view centered on
$\eta$ Carinae, part of the Keyhole Nebula, and Tr~14.  The new
contiguous mosaic image of the bright region of the nebula was
obtained with the ACS/WFC camera using the H$\alpha$ (F658N) filter.
It has been released as a stunning color image as part of the Hubble
Heritage project\footnote{{\tt
http://heritage.stsci.edu/2007/16/index.html}}, the main part of which
is shown in Figure~\ref{fig:heritage}.

The huge HST mosaic image is one of the largest made by HST, and
contains an enormous amount of detail.  A small taste of the imminent
results can be seen in Figure~\ref{fig:jets}, which displays just a
small few arcminute-wide section of the nebula.  Here we see a dust
pillar just north of the Tr~14 cluster which is being eroded by stellar
winds.  The zoomed-in view shows two spectacular bipolar Herbig-Haro
jets, HH~901 and 902, emerging from near the tips of
two smaller dust
pillars.  Several other HH jets are seen in the HST data as well.  In
addition to the first HH jet to be discovered in Carina (HH~666; Smith
et al.\ 2004$a$) these HH jets are signposts of active ongoing star
formation still embedded within the dust pillars that are now being
demolished.

The new HST data provide an immensely complex and large dataset, and
are unfortunatley still in the early stages of analysis, so we cannot
provide a summary of the results (such as a Table of all
the HH jets discovered), but combined with data from Spitzer, we
expect them to shed important new light on the outflow activity and
the embedded stellar population in Carina.  Such papers are currently
in preparation (Smith et al.).

\begin{figure}[!ht]
\centering
\includegraphics[width=\textwidth,draft=False]{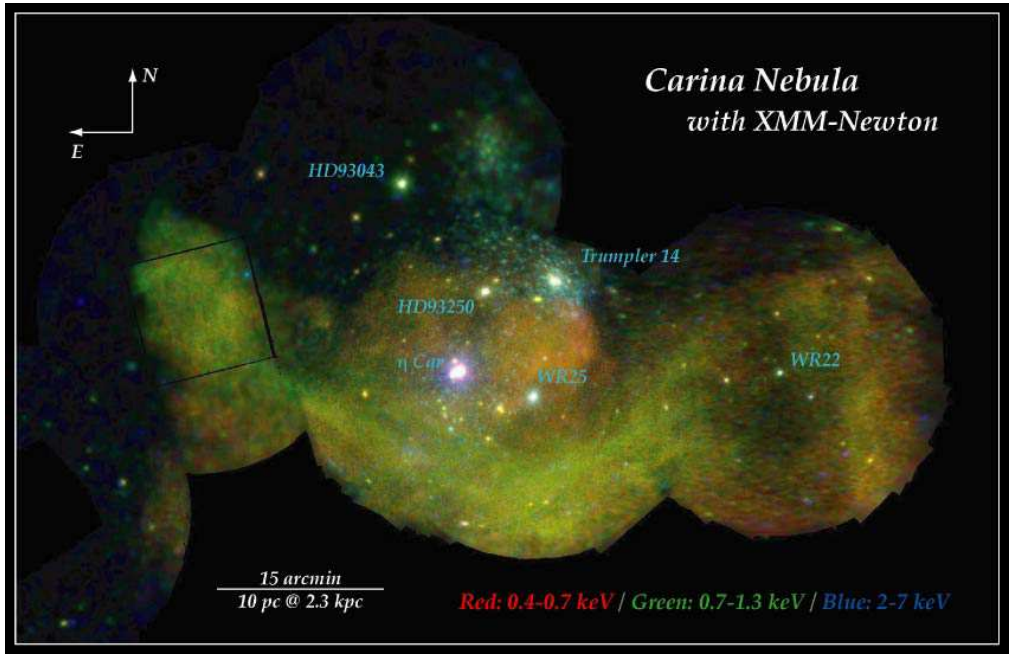}
\caption{Wide-field X-ray image of the inner Carina Nebula taken with
  XMM.  Soft X-rays (0.4-0.7 keV) are in red, medium/soft X-rays
  (0.7-1.3 keV) are in green, and harder X-rays are in blue.  Several
  key sources are identified.  {\it Image credit}: K.\ Hamaguchi,
  NASA, ESA}
\label{fig:xmm}
\end{figure}

\subsection{Recent and Planned X-ray Observations with {\it XMM} and
  {\it Chandra}}

The diffuse X-ray emission in the Carina Nebula is unusually luminous,
about 10-100 times stronger than other giant H~{\sc ii} regions, and
it was the first among giant H~{\sc ii} regions to be discovered to
have diffuse X-rays (Seward et al.\ 1979; Seward \& Chlebowski 1982).
Not surprisingly then, it is an important target for the
currently-available suite of X-ray telescopes in space such as
XMM-Newton, the Chandra X-ray Observatory, and the Suzaku (Astro-E2)
mission.

Figure~\ref{fig:xmm} shows a wide-field mosaic image of the diffuse
X-ray emission in Carina (K.\ Hamaguchi, private comm.), composed of
several individual pointings taken with the XMM-Newton satellite.  It
illustrates that the emission from Carina is a complex combination of
point sources from O stars, WNH stars, and colliding wind binaries,
plus diffuse soft X-ray emission that spans more than 1$^{\circ}$.
Many of the point sources are fascinating systems in their own right
(many are colliding-wind binaries like $\eta$ Car and WR25; see
Corcoran 2005, Raassen et al.\ 2003), but we will not review the X-ray
properties of these massive stars here.  Some fraction of the diffuse
emission may also be contributed by unresolved low-mass T Tauri stars
that have formed alongside the massive members of these clusters and
throughout the nebula, so spatial resolution is an issue when
interpreting the data.  With the superior spatial resolution of
Chandra, preliminary studies of small regions around Tr~16 and Tr~14
have revealed hundreds of point sources (Evans et al.\ 2003; Townsley
2006; Sanchawala et al.\ 2007), but so far only a portion of the inner
nebula has been studied with Chandra. An accurate census of the
low-mass X-ray emitting stellar population will be an important goal
of an upcoming large survey of Carina with Chandra (PI: L.\ Townsley).

For the time being, one of the most pressing questions regarding X-ray
observations of Carina is the physical origin of the unusually bright
diffuse X-ray emission.  The first explanation that usually comes to
mind is that there may have been a recent supernova explosion in
Carina.  If that were true, it would have stunning implications from
the point of view of stellar evolution.  Namely, if a star exploded in
the vicinity of Tr~16 and Tr~14, we might expect that this progenitor
star had an initial mass larger than that of $\eta$ Carinae because it
reached the end of its life sooner.  Little is known about the end
fates of stars that massive, and it might alter our interpretation of
$\eta$ Car's role in the history of the Carina region.  It would also
mean that Carina is an example of a star-forming region caught in the
act of polluting protostellar disks with supernova ejecta, with
far-reaching implications for the origin of our own Solar System (e.g.,
Tachibana \& Huss 2003; Hester et al.\ 2004; Desch \& Ouellete 2005).
Thus, alternatives should be considered carefully.

The diffuse emission cannot be accounted for fully by an extrapolation
of the discrete source luminosity function, but on the other hand,
Seward \& Chlebowsky (1982) argued that energy input to the diffuse
plasma from collisions of the many O-star and WNH-star winds with the
ambient medium was sufficient (see also Townsley et al.\ 2003; Ezoe et
al.\ 2006).  Smith (2006$a$) also showed that the integrated
mechanical energy input from stellar winds was more than enough to
power the expansion of the region (see also Smith \&
Brooks 2007).

Thus, there appears to be no {\it need} to invoke a recent supernova
explosion on energetic grounds alone, but there are some other clues
that make the specter of a supernova a hard hypothesis to rule out.
For instance, the spectral energy distribution of the diffuse X-ray
emission in Carina varies with position in a peculiar way.  Studies of
the diffuse emission with XMM-Newton, Chandra, and Suzaku (Hamaguchi
et al.\ 2007; Townsley 2006) show that the diffuse emission
immediately south of Tr~14 and north of WR25 (reddish hue in
Fig.~\ref{fig:xmm}) has both an intermediate/hard component at 0.62
keV and a softer component at 0.26 keV, with relatively normal
abundances.  By contrast the emission further south (greenish hue
running east/west across the nebula in Fig.~\ref{fig:xmm}) requires
only the 0.6 keV component (no softer X-rays) and seems to show
enhanced Ne and Fe abundances.  Hamaguchi et al.\ (2007) suggest that
the lack of strong nitrogen enrichment and normal oxygen abundances
favor the SN hypothesis, because one would expect stellar winds to
contribute more nitrogen rich material.  However, Smith \& Morse
(2004) showed that even in the case of an evolved massive star like
$\eta$ Car, the N-rich ejecta are tightly confined to within about
1\arcmin\ of the star, with progressively more normal N/O ratios at
larger radii from the star.  Also, examining the different color of
X-ray emission in Figure~\ref{fig:xmm} that represents different
energies, one might wonder why the plasma that lacks a soft component
and has enhanced Fe and Ne abundances (green) is spatially coincident
with the V-shaped dust lane that bisects the nebula (see
Fig.~\ref{fig:quad}).  This region has the highest line-of-sight
column density in the nebula (Smith \& Brooks 2007), so it is not
clear how strong absorption through a clumpy medium might affect the
spectra.  The alternative that a supernova focussed Fe-rich ejecta
exclusively toward the dust lane seems dubious.  These questions do
not appear to have easy solutions with available data.

Overall, we still do not know the origin of the diffuse X-ray emission
in Carina, and we look forward with great anticipation to the results
of the upcoming large survey with Chandra.

\section{Closing Remarks}

In this review, we have highlighted recent work that portrays the
Carina Nebula as a rich star-forming region, and a laboratory in which
to study feedback from massive stars.  It gives examples of the same
detailed phenomena seen in closer regions like Orion, while sampling a
much more extreme stellar population analogous to that of 30 Dor.  The
vicinity of the Keyhole Nebula gives us a view of the remnant of a GMC
that has already been dispersed by massive stars in Tr~16, while the
northern cloud near the younger Tr~14 cluster is more intact, still
awaiting vigorous star formation.  The South Pillars currently harbor
the most active star formation in Carina, where stellar winds and
radiation are sculpting dust pillars and probably triggering a new
generation of stars.

Despite these recent highlights, we suspect that the most exciting
work on Carina remains to be done, with more comprehensive study of
the overall properties of the gas and embedded stars.  Recent surveys
with {\it HST}, {\it Spitzer}, and {\it Chandra} will soon yield a
rich harvest of new results.  The brightness and complexity of Carina
will undoubtedly provide new discoveries and exciting new challenges
for future IR observatories like SOFIA and Herschel, and it will
no-doubt be a bonanza for planned millimeter projects in the southern
hemisphere like ALMA and the single-dish telescopes at the same site.\\

{\bf Acknowledgements.} \ We thank Nolan Walborn for helpful comments
after a careful reading of this manuscript as the referee, as well as
D.\ Malin, J.\ Rathborne, and Y.\ Yonekura for use of their data.  We
also thank the ATNF Distinguished Visitor Program, providing us with
an opportunity to collaborate on this review in person in Australia.
N.S.\ was supported by NASA through grants GO-10241 and GO-10475 from
STScI.


\begin{thebibliography}{}

\bibitem[\protect\citeauthoryear{Allen}{Allen}{1979}]{Allen79}  Allen,
  D.A.\ 1979, MNRAS, 189, 1

\bibitem[\protect\citeauthoryear{Allen \& Hillier}{Allen \&
    Hillier}{1993}]{Allen93} Allen, D.A. \& Hillier, D.J.\ 1993,
    PASA, 10, 338

\bibitem[\protect\citeauthoryear{Ascenso et al.}{2007}]{ascenso07}
    Ascenso, J., Alves, J., Vicente, S., \& Lago, M.T.V.T.\ 2007,
    A\&A, 976, 199

\bibitem[\protect\citeauthoryear{Blaauw}{Blaauw}{1964}]{Blaauw64}
  Blaauw, A.\ 1964, ARAA, 2, 213

\bibitem[\protect\citeauthoryear{Bok}{Bok}{1932}]{Bok32} Bok, B.J.\
  1932, Harvard Reprints, 1, 77

\bibitem[\protect\citeauthoryear{Brooks et~al.}{Brooks
  et~al.}{2000}]{Brooks00} Brooks, K.J., Burton, M.G., Rathborne,
  J.M. et al.\ 2000, MNRAS, 319, 95

\bibitem[\protect\citeauthoryear{Brooks et~al.}{Brooks
  et~al.}{2003}]{Brooks03} Brooks, K.J., Cox, P., Schneider, N.,
  et~al., 2003, A\&A, 412, 751

\bibitem[\protect\citeauthoryear{Brooks et~al.}{Brooks
  et~al.}{2005}]{Brooks05} Brooks, K.J., Garay, G., Nielbock, M.,
  Smith, N., \& Cox P.\ 2005, ApJ, 634, 436

\bibitem[\protect\citeauthoryear{Brooks, Storey, \& Whiteoak}{Brooks
  et~al.}{2001}]{Brooks01} Brooks, K.J., Storey, J.W.V., \& Whiteoak,
  J.B.\ 2001, MNRAS, 327, 46

\bibitem[\protect\citeauthoryear{Brooks, Whiteoak, \& Storey}{Brooks
  et~al.}{1998}]{Brooks98} Brooks, K.J., Whiteoak, J.B., \& Storey,
  J.W.V.\ 1998, Pub.\ Astron.\ Soc.\ Australia, 15(2), 202

\bibitem[\protect\citeauthoryear{Carraro}{Carraro}{2002}]{car02}
Carraro, G.\ 2002, MNRAS, 331, 785

\bibitem[\protect\citeauthoryear{Carraro \& Patat}{Carraro \&
Patat}{2001}]{car01a} Carraro, G.\ \& Patat, F.\ 2001, A\&A, 379, 136

\bibitem[\protect\citeauthoryear{Carraro et al.}{Carraro et
  al.}{2001}]{car01b} Carraro, G., Patat, F., \& Baumgardt, H.\ 2001,
  A\&A, 371, 107

\bibitem[\protect\citeauthoryear{Carraro et al.}{Carraro et
  al.}{2004}]{car04} Carraro, G., Romaniello, M., Ventura, P., \&
  Patat, F.\ 2004, A\&A, 418, 525

\bibitem[\protect\citeauthoryear{Caswell}{Caswell}{1998}]{Caswell98}
  Caswell, J.L.\ 1998, MNRAS, 297, 215

\bibitem[\protect\citeauthoryear{Caswell \& Haynes}{Caswell \&
  Haynes}{1975}]{Caswell75} Caswell, J.L. \& Haynes, R.F.\ 1975,
  MNRAS, 173, 649

\bibitem[\protect\citeauthoryear{Cersosimo, Azcarate, \&
  Colomb}{Cersosimo et~al.}{1984}]{Cersosimo84} Cersosimo, J.C.,
  Azcarate, I.N., \& Colomb, F.R.\ 1984, ApJ, 24, 1

\bibitem[\protect\citeauthoryear{Chan \& Onaka 2000}{Chan \&
    Onaka}{2000}]{ChanO} Chan, K.W. \& Onaka, T.\ 2000, ApJ, 533, L33

\bibitem[\protect\citeauthoryear{Cor}{Cor}{2005}]{Cor05} Corcoran,
  M.F.\ 2005, AJ, 129, 2018

\bibitem[\protect\citeauthoryear{Cox}{Cox}{1995}]{Cox952}
  Cox, P.\ 1995, Rev.\ Mex.\ Astron.\ Astrofis.\ Ser.\ Conf., 2, 105

\bibitem[\protect\citeauthoryear{Cox \& Bronfman}{Cox \&
  Bronfman}{1995}]{Cox951} Cox, P. \& Bronfman, L.\ 1995, A\&A, 299,
  583

\bibitem[\protect\citeauthoryear{Crowther et al.}{Crowther et
    al.}{1995}]{crowther} Crowther, P.A., Smith, L.J., Hillier, D.J.,
    \& Schmutz, W.\ 1995, A\&A, 293, 427

\bibitem[\protect\citeauthoryear{Davidson \& Humphreys}{Davidson \&
Humphreys}{1997}]{dh97} Davidson, K., \& Humphreys, R.M.\ 1997,
ARA\&A, 35, 1

\bibitem[\protect\citeauthoryear{Damineli et al.}{Damineli et
al.}{2000}]{dam00} Damineli, A., Kaufer, A., Wolf, B. et al.\ 2000,
ApJ, 528, L101

\bibitem[\protect\citeauthoryear{de~Graauw et~al.}{de~Graauw
  et~al.}{1981}]{deGraauw81} de~Graauw, T., Lidholm, S., Fitton, B.,
  et~al.\ 1981, A\&A, 102, 257

\bibitem[\protect\citeauthoryear{DeGioia-Eastwood et
    al.}{DeGioia-Eastwood et al.}{2001}]{degioia} DeGioia-Eastwood,
    K., Throop, H., Walker, G., \& Cudworth, K.M.\ 2001, ApJ, 549, 578

\bibitem[\protect\citeauthoryear{Deharveng \& Maucherat}{Deharveng \&
  Maucherat}{1975}]{Deharveng75} Deharveng, L. \& Maucherat, M.\ 1975,
  A\&A, 41, 27

\bibitem[\protect\citeauthoryear{Desch \& Ouellete}{Desch \&
Ouellete}{2005}]{DO05} Desch, S.J. \& Ouellete, N.\ 2005, Lunar
  Planet Sci.\ Conf., 36, 1327

\bibitem[\protect\citeauthoryear{Dickel}{Dickel}{1974}]{Dickel741}
  Dickel, H.R.\ 1974, A\&A, 31, 11

\bibitem[\protect\citeauthoryear{Dickel \& Wall}{Dickel \&
  Wall}{1974}]{Dickel742} Dickel, H.R. \& Wall, J.V.\ 1974, A\&A, 31,
  5

\bibitem[\protect\citeauthoryear{Elliot}{Elliot}{1979}]{Elliot79}
  Elliot, K.H.\ 1979, MNRAS, 186, 9

\bibitem[\protect\citeauthoryear{Evans et al.}{Evans et
  al.}{2003}]{Evans03} Evans, N.R., Seward, F.D., Krauss, M.I. et
  al.\ 2003, ApJ, 589, 509

\bibitem[\protect\citeauthoryear{Ezoe et al.}{Ezoe et
  al.}{2006}]{Ezoe06} Ezoe, Y., Kokubun, M., Makishima, K. et al.\
  2006, ApJ, 638, 860

\bibitem[\protect\citeauthoryear{Feinstein}{Feinstein}{1982}]
  {Feinstein82} Feinstein,A.\ 1982, AJ, 87, 1012

\bibitem[\protect\citeauthoryear{Feinstein}{Feinstein}{1995}]
  {Feinstein95} Feinstein,A.\ 1995, RevMexAA, Ser.\ Conf., 2, 57

\bibitem[\protect\citeauthoryear{Feinstein \& Marraco}{Feinstein \&
  Marraco}{1980}]{Feinstein80} Feinstein, A. \& Marraco, H.~G.\ 1980,
  PASP, 92, 266

\bibitem[\protect\citeauthoryear{Feinstein, Marraco, \&
  Muzzio}{Feinstein et~al.}{1973}]{Feinstein73} Feinstein, A., Marraco,
  H.~G., \& Muzzio, J.~C.\ 1973, A\&AS, 12, 331

\bibitem[\protect\citeauthoryear{Feinstein, Moffat, \&
  Fitzgerald}{Feinstein et~al.}{1980}]{Feinstein80} Feinstein, A.,
  Moffat, A.F.J., \& Fitzgerald, M.P.\ 1980, AJ, 85, 708

\bibitem[\protect\citeauthoryear{Figer}{Figer}{1999}]{Figer99} Figer,
  D.F., Kim, S.S., Morris, M. et al.\ 1999, ApJ, 525, 750

\bibitem[\protect\citeauthoryear{Figer}{Figer}{2002}]{Figer02} Figer,
  D.F., Najarro, F., Gilmore, D. et al.\ 2002, ApJ, 581, 258

\bibitem[\protect\citeauthoryear{Fitzgerald \& Mehta}{Fitzgerald \&
Mehta}{1987}]{fm87} Fitzgerald, M.P.\ \& Mehta, S.\ 1987, MNRAS, 228,
545

\bibitem[\protect\citeauthoryear{Forte}{Forte}{1978}]{forte} Forte,
  J.C.\ 1978, AJ, 83, 1199

\bibitem[\protect\citeauthoryear{Frew}{Frew}{2004}]{Frew04} Frew,
  D.J.\ 2004, J.\ of Astron.\ Data, 10, 6

\bibitem[\protect\citeauthoryear{Gaensler et al.}{Gaensler et
    al.}{2005}]{gaensler} Gaensler, B.M., McClure-Griffiths, N.M.,
    Oey, M.S., Haverkorn, M., Dickey, J.M., \& Green, A.J.\ 2005, ApJ,
    620, L95

\bibitem[\protect\citeauthoryear{Garcia \& Walborn}{Garcia \&
Walborn}{2000}]{gw00} Garcia, B. \& Walborn, N.R.\ 2000, PASP, 112,
1549

\bibitem[\protect\citeauthoryear{Garcia et al.}{Garcia et
al.}{1998}]{gar98} Garcia, B., Malaroda, S., Levato, H., Morrell, N.,
\& Grosso, M.\ 1998, PASP, 110, 53

\bibitem[\protect\citeauthoryear{Gardner, Dickel, \& Whiteoak}{Gardner
  et~al.}{1973}]{Gardner73} Gardner, F.F., Dickel, H.R., \& Whiteoak,
  J.B.\ 1973, A\&A, 23, 51

\bibitem[\protect\citeauthoryear{Gardner \& Morimoto}{Gardner \&
  Morimoto}{1968}]{Gardner68} Gardner, F.F. \& Morimoto, M.\ 1968,
  Aust.\ J.\ Phys., 21, 881

\bibitem[\protect\citeauthoryear{Gardner et al.}{Gardner
  et~al.}{1970}]{Gardner70} Gardner, F.F., Milne, D.K., Mezger, P.G.,
  \& Wilson, T.L.\ 1970, A\&A, 7, 349

\bibitem[\protect\citeauthoryear{Gaviola}{Gaviola}{1950}]{Gaviola50}
Gaviola, E.\ 1950, ApJ, 111, 408

\bibitem[\protect\citeauthoryear{Ghosh et~al.}{Ghosh
et~al.}{1988}]{Ghosh88} Ghosh, S.K., Iyengar, K.V.K., Rengarajan,
S.N., et~al.\ 1988, ApJ, 330, 928

\bibitem[\protect\citeauthoryear{Grabelsky et~al.}{Grabelsky
  et~al.}{1988}]{Grabelsky88} Grabelsky, D.A., Cohen, R.S., Bronfman,
  L., \& Thaddeus, P.\ 1988, ApJ, 331, 181

\bibitem[\protect\citeauthoryear{Habing}{Habing}{1968}]{Habing68}
  Habing, H.J.\ 1968, Bull.\ Astron.\ Neth., 19, 421

\bibitem[\protect\citeauthoryear{Haisch}{Haisch}{2000}]{Haisch00}
  Haisch, K.E., Lada, E.A., \& Lada, C.J.\ 2000, AJ, 120, 1396

\bibitem[\protect\citeauthoryear{H\"{a}gele et al.}{H\"{a}gele et
    al.}{2004}]{Hagele04} H\"{a}gele, G.F., Albacete Colombo, J.F.,
    Barba, R.H., \& Bosch, G.L.\ 2004, MNRAS, 355, 1237

\bibitem[\protect\citeauthoryear{Harvey, Hoffmann, \& Campbell}{Harvey
 et~al.}{1979}]{Harvey79} Harvey, P.M., Hoffmann, W.F., \& Campbell,
 M.F.\ 1979, ApJ, 227, 114

\bibitem[\protect\citeauthoryear{Herbst}{Herbst}{1976}]{herbst} Herbst,
  W.\ 1976, ApJ, 208, 923

\bibitem[\protect\citeauthoryear{Herschel}{Herschel}{1847}]{Herschel47}
  Herschel, J.F.W.\ 1847, {\it Results of Observations Made During the
  Years 1834, 5, 6, 7, 8 at the Cape of Good Hope} (London: Smith,
  Elder)

\bibitem[\protect\citeauthoryear{Hester et al.}{Hester et
  al.}{2004}]{Hes04} Hester, J.J., Desch, S.J., Healy, K.R., \&
  Leshin, L.A.\ 2004, Science, 304, 1116

\bibitem[\protect\citeauthoryear{Hollenbach \& Tielens}{Hollenbach \&
   Tielens}{1997}]{Hollenbach97}
Hollenbach, D.J. \& Tielens, A.G.G.M.\ 1997, ARA\&A, 35, 179

\bibitem[\protect\citeauthoryear{Huchtmeier \& Day}{Huchtmeier \&
  Day}{1975}]{Huchtmeier75} Huchtmeier, W.K. \& Day G.A.\ 1975,
  A\&A, 41, 153

\bibitem[\protect\citeauthoryear{Humphreys}{Humphreys}{1978}]{hum}
Humphreys, R.M.\ 1978, ApJS, 38, 309

\bibitem[\protect\citeauthoryear{Israel et~al.}{Israel
   et~al.}{1996}]{Israel96} Israel, F.P., Maloney, P.R., Geis, N.,
   et~al.\ 1996, ApJ, 465, 738

\bibitem[\protect\citeauthoryear{Jones}{Jones}{1973}]{Jones73} Jones,
  B.B.\ 1973, Aust. J. Phys, 26, 545

\bibitem[\protect\citeauthoryear{Kaltcheva \& Georgiev}{Kaltcheva \&
Georgiev}{1993}]{kg93} Kaltcheva, N.T.\ \& Georgiev, L.N.\ 1993,
MNRAS, 261, 847

\bibitem[\protect\citeauthoryear{Lada \& Lada}{Lada \&
    Lada}{2003}]{LL03} Lada, C.J. \& Lada, E.A.\ 2003, ARAA, 41, 57

\bibitem[\protect\citeauthoryear{Laurent, Paul, \& Pettini}{Laurent
  et~al.}{1982}]{Laurent82} Laurent, C., Paul, J.A., \& Pettini, M.\
  1982, ApJ, 260, 163

\bibitem[\protect\citeauthoryear{Lee et al.}{Lee et al.}{2000}]{lee}
  Lee, D.H., Min, K.W., Dixon, W.V.D. et al.\ 2000, ApJ, 545, 885

\bibitem[\protect\citeauthoryear{Levato et al.}{Levato et
  al.}{1991}]{lev} Levato, H., Malaroda, S., Morrell, N., Garcia, B.,
  \& Hernandez, C.\ 1991, ApJS, 75, 869

\bibitem[\protect\citeauthoryear{Lopez \& Meaburn}{Lopez \&
  Meaburn}{1986}]{Lopez86} Lopez, J.A. \& Meaburn, J.\ 1986, Rev.\
  Mex.\ Astron.\ Astrofis., 13, 27

\bibitem[\protect\citeauthoryear{Marraco et al.}{Marraco et
al.}{1993}]{mar93} Marraco, H.G., Vega, E.I., \& Vrba, F.J.\ 1993, AJ,
105, 258

\bibitem[\protect\citeauthoryear{Massey \& Johnson}{Massey \&
    Johnson}{1993}]{MJ93} Massey, P. \& Johnson, J.\ 1993, AJ, 105,
    980

\bibitem[\protect\citeauthoryear{Massey \& Hunter}{Massey \&
    Hunter}{1998}]{MH98} Massey, P. \& Hunter, D.A.\ 1998, ApJ, 493, 180

\bibitem[\protect\citeauthoryear{Meaburn, Lopez, \& Keir}{Meaburn
  et~al.}{1984}]{Meaburn84} Meaburn, J., Lopez, J.A., \& Keir, D.\ 1984,
  MNRAS, 211, 267

\bibitem[\protect\citeauthoryear{Megeath et al.}{Megeath et
    al.}{1996}]{Megeath96} Megeath, S.T., Cox, P., Bronfman, L., \&
    Roelfsema, P.R.\ 1996, A\&A, 305, 296

\bibitem[\protect\citeauthoryear{Mizutani et al.}{Mizutani et
    al.}{2004}]{Mizutani04} Mizutani, M., Onaka, T., \& Shibai, H.\
    2004, A\&A, 423, 579

\bibitem[\protect\citeauthoryear{Mahhat \& Vogt}{Moffat \&
Vogt}{1975}]{} Moffat, A.F.J. \& Vogt, N.\ 1975, A\&A, 20, 125

\bibitem[\protect\citeauthoryear{Moffat et al.}{Moffat et
    al.}{2002}]{Moffat02} Moffat, A.F.J., Corcoran, M. F., Stevens,
    I.R. et al.\ 2002, ApJ, 573, 191

\bibitem[\protect\citeauthoryear{Morrell et al.}{Morrell et
al.}{1988}]{mor} Morrell, N., Garcia, B., \& Levato, H.\ 1988, PASP,
100, 1431

\bibitem[\protect\citeauthoryear{Najarro et al.}{Najarro et
  al.}{2004}]{Najarro04} Najarro, F., Figer, D.F., Hillier, D.J., \&
  Kudritzki, R.P.\ 2004, ApJ, 611, L105

\bibitem[\protect\citeauthoryear{O'Dell}{O'Dell}{2001}]{ODell01}
O'Dell, C.~R.\ 2001, ARA\&A, 39, 99

\bibitem[\protect\citeauthoryear{Raassen et al.}{Raassen et
    al.}{2003}]{raassen} Raassen, A.J.J., van der Hucht, K.A. Mewe,
    R. et al.\ 2003, A\&A, 402, 653

\bibitem[\protect\citeauthoryear{Rathborne et~al.}{Rathborne
  et~al.}{2004}]{Rathborne04} Rathborne, J.M., Brooks, K.J., Burton,
  M.G., Cohen, M., \& Bontemps, S.\ 2004, A\&A, 418, 563

\bibitem[\protect\citeauthoryear{Rathborne et~al.}{Rathborne
  et~al.}{2002}]{Rathborne02} Rathborne, J.M., Burton, M.~G., Brooks,
  K.J., et~al.\ 2002, MNRAS, 331, 85

\bibitem[\protect\citeauthoryear{Reipurth et al.}{Reipurth et
    al.}{2003}]{reipurth03} Reipurth, B., Raga, A., \& Heathcote, S.\
    2003, AJ, 126, 1925

\bibitem[\protect\citeauthoryear{Retallack}{Retallack}{1983}]{Retallack83}
Retallack, D.~S.\ 1983, MNRAS, 204, 669

\bibitem[\protect\citeauthoryear{Sanchawala et al.}{Sanchawala et
    al.}{2007}]{sanchawala} Sanchawala, K., Chen, W.P., Lee, H.T.,
    al.\ 2007, ApJ, 656, 462

\bibitem[\protect\citeauthoryear{Schwartz}{Schwartz}{1990}]{Scwartz90}
  Schwartz, R.D., Persson, S.E., \& Hamann, F.W.\ 1990, AJ, 100, 793

\bibitem[\protect\citeauthoryear{Seward \& Chlebowski}{Seward \&
    Chlebowski}{1982}]{seward82} Seward, F.D. \& Chlebowski, T.\
    1982, ApJ, 256, 530

\bibitem[\protect\citeauthoryear{Seward et al.}{Seward et
    al.}{1979}]{seward79} Seward, F.D., Forman, W.R., Giacconi, R., et
    al.\ 1979, ApJ, 234, L55

\bibitem[\protect\citeauthoryear{Shaver \& Goss}{Shaver \&
  Goss}{1970}]{Shaver70} Shaver, P.~A. \& Goss, W.~M.\ 1970, Australian
  J. Phys. Suppl., 14, 133

\bibitem[\protect\citeauthoryear{Sher}{Sher}{1965}]{sher} Sher, D.\
1965, Quart.\ J.\ R.\ Astron.\ Soc., 6, 299

\bibitem[\protect\citeauthoryear{Shobbrook}{Shobbrook}{1980}]{sho80}
Shobbrook, R.R.\ 1980, MNRAS, 192, 821

\bibitem[\protect\citeauthoryear{Shobbrook \& Lynga}{Shobbrook \&
Lynga}{1994}]{sho94} Shobbrook, R.R. \& Lynga, G. 1994, MNRAS, 269,
857

\bibitem[\protect\citeauthoryear{Smith}{Smith}{2002a}]{Smith02a}
Smith, N.\ 2002a, MNRAS, 337, 1252

\bibitem[\protect\citeauthoryear{Smith}{Smith}{2002b}]{Smith02b}
Smith, N.\ 2002b, MNRAS, 331, 7

\bibitem[\protect\citeauthoryear{Smith}{Smith}{2006a}]{Smith06}
Smith, N.\ 2006a, MNRAS, 367, 763

\bibitem[\protect\citeauthoryear{Smith}{Smith}{2006b}]{Smith06b}
Smith, N.\ 2006b, ApJ, 644, 1151

\bibitem[\protect\citeauthoryear{Smith \& Brooks}{Smith \&
Brooks}{2007}]{SB07} Smith, N. \& Brooks, K.J.\ 2007, MNRAS, 379,
1279

\bibitem[\protect\citeauthoryear{Smith \& Conti}{Smith \&
Conti}{2008}]{SC08} Smith, N. \& Conti, P.S.\ 2008, ApJ, 679, 1467

\bibitem[\protect\citeauthoryear{Smith \& Morse}{Smith \&
Morse}{2004}]{SM04} Smith, N. \& Morse, J.A.\ 2004, ApJ, 605, 854

\bibitem[\protect\citeauthoryear{Smith et~al.}{Smith
et~al.}{2000}]{Smith00} Smith, N., Egan, M.P., Carey, S., et~al.\
2000, ApJ, 532, L145

\bibitem[\protect\citeauthoryear{Smith, Bally, \& Morse}{Smith
et~al.}{2003a}]{Smith03a} Smith, N., Bally, J., \& Morse, J.A.\ 2003a,
ApJ, 587, L105

\bibitem[\protect\citeauthoryear{Smith et al.}{Smith et
al.}{2003b}]{Smith03b} Smith, N., Gehrz, R.D., Hinz, P.M. et al.\
2003b, AJ, 125, 1458

\bibitem[\protect\citeauthoryear{Smith, Bally, \& Brooks}{Smith
et~al.}{2004a}]{Smith004a} Smith, N., Bally, J., \& Brooks, K.J.\
2004a, AJ, 127, 2793

\bibitem[\protect\citeauthoryear{Smith, Barba, \& Walborn}{Smith
et~al.}{2004b}]{Smith04b} Smith, N., Barba, R.H., \& Walborn, N.R.\
2004b, MNRAS, 351, 1457

\bibitem[\protect\citeauthoryear{Smith, Morse, \& Bally}{Smith
et~al.}{2005a}]{Smith05a} Smith, N., Morse, J.A., \& Bally, J.\ 2005a,
AJ, 130, 1778

\bibitem[\protect\citeauthoryear{Smith, Stassun, \& Bally}{Smith
et~al.}{2005b}]{Smith05b} Smith, N., Stassun, K.G., \& Bally, J.\ 2005b,
AJ, 129, 888

\bibitem[\protect\citeauthoryear{Smith, Bally, \& Walawender}{Smith
et~al.}{2007}]{Smith07} Smith, N., Bally, J., \& Walawender, J.\ 2007,
  AJ, 134, 846

\bibitem[\protect\citeauthoryear{Smith}{Smith}{1987}]{Smith87} Smith,
R.G.\ 1987, MNRAS, 227, 943

\bibitem[\protect\citeauthoryear{Tachibana \& Huss}{Tachibana \&
    Huss}{2004}]{TH} Tachibana, S. \& Huss, G.R.\ 2004, ApJ, 588, L41

\bibitem[\protect\citeauthoryear{Tapia et al.}{Tapia et
    al.}{1988}]{Tapia88} Tapia, M., Roth, M., Marraco, H., \& Ruiz,
    M.T..\ 1988, MNRAS, 232, 661

\bibitem[\protect\citeauthoryear{Tapia}{Tapia}{1995}]{Tapia95} Tapia,
M.\ 1995, Rev.\ Mex.\ Astron.\ Astrofis.\ Ser.\ Conf., 2, 87

\bibitem[\protect\citeauthoryear{Tapia et al.}{Tapia et
    al.}{2003}]{Tapia03} Tapia, M., Roth, M., Vazquez, R.A., \&
    Feinstein A.\ 2003, MNRAS, 339, 44

\bibitem[\protect\citeauthoryear{Tapia}{Tapia}{2004}]{Tapia04} Tapia,
M.\ 2004, Rev.\ Mex.\ Astron.\ Astrofis.\ Ser.\ Conf., 22, 73

\bibitem[\protect\citeauthoryear{Tapia et al.}{Tapia et
    al.}{2006}]{Tapia06} Tapia, M., Persi, P., Bohigas, J., Roth, M.,
    \& Gomez, M.\ 2006, MNRAS, 367, 513

\bibitem[\protect\citeauthoryear{Tateyama, Strauss, \&
  Kaufmann}{Tateyama et~al.}{1991}]{Tateyama91} Tateyama, C.E.,
  Strauss, F.M., \& Kaufmann, P.\ 1991, MNRAS, 249, 716

\bibitem[\protect\citeauthoryear{Thackeray}{Thackeray}{1950}]{Thackeray50a}
  Thackeray, A.D.\ 1950a, 110, 529

\bibitem[\protect\citeauthoryear{Thackeray}{Thackeray}{1950}]{Thackeray50b}
  Thackeray, A.D.\ 1950b, MNRAS, 110, 524

\bibitem[\protect\citeauthoryear{Th\'{e} et al.}{Th\'{e} et
    al.}{1980}]{the80} Th\'{e}, P.S., Bakker, R., \& Tjin A Djie,
    H.R.E.\ 1980, A\&A, 89, 209

\bibitem[\protect\citeauthoryear{Th\'{e} \& Vleeming}{Th\'{e} \&
    Vleeming}{1971}]{the71} Th\'{e}, P.S. \& Vleeming, G.\ 1971,
    A\&A, 14, 120

\bibitem[\protect\citeauthoryear{Townsley}{Townsley}{2006}]{Town06}
Townsley, L.K.\ 2006, in Proc.\ STScI May Symposium, {\it Massive
stars: From PopIII and GRBs to the Milky Way}, ed.\ M.\ Livio
(astro-ph/0608173)

\bibitem[\protect\citeauthoryear{Townsley et al.}{Townsley et
al.}{2003}]{Town03} Townsley, L.K., Feigelson, E.D., Montmerle, T.,
Broos, P.S., Chu, Y.H., \& Garmire, G.P.\ 2003, ApJ, 593, 874

\bibitem[\protect\citeauthoryear{Turner et al.}{Turner et
  al.}{1980}]{tur80} Turner, D.G., Grieve, G.R., Herbst, W., \&
  Harris, W.E.\ 1980, AJ, 85, 1193

\bibitem[\protect\citeauthoryear{Vazquez et al.}{Vazquez et
al.}{1996}]{vaz06} Vazquez, R.A., Baume, G., Feinstein, A., \& Prado,
P.\ 1996, A\&AS, 116, 75

\bibitem[\protect\citeauthoryear{Walborn}{Walborn}{1973}]{Walborn73}
 Walborn, N.R.\ 1973, ApJ, 179, 517

\bibitem[\protect\citeauthoryear{Walborn}{Walborn}{1975}]{Walborn75}
 Walborn, N.R.\ 1975, ApJ, 202, L129

\bibitem[\protect\citeauthoryear{Walborn}{Walborn}{1982}]{Walborn82b}
 Walborn, N.R.\ 1982, ApJS, 48, 145

\bibitem[\protect\citeauthoryear{Walborn}{Walborn}{1995}]{Walborn95}
Walborn, N.R.\ 1995, Rev.\ Mex.\ Astron.\ Astrofis.\ Ser.\ Conf., 2, 51

\bibitem[\protect\citeauthoryear{Walborn}{Walborn}{2002}]{Walborn02}
 Walborn, N.R.\ 2002, ASP Conf.\ Ser.\ 267, {\it Hot Star Workshop
 III: The Earliest Stages of Massive Star Birth}, ed.\ P.A.\ Crowther
 (San Francisco: ASP), 111

\bibitem[\protect\citeauthoryear{Walborn \& Hesser}{Walborn \&
  Hesser}{1975}]{Walborn75} Walborn, N.R. \& Hesser, J.E.\ 1975, ApJ,
  199, 535

\bibitem[\protect\citeauthoryear{Walborn \& Hesser}{Walborn \&
  Hesser}{1982}]{Walborn82} Walborn, N.R. \& Hesser, J.E.\ 1982, ApJ,
  252, 156

\bibitem[\protect\citeauthoryear{Walborn \& Liller}{Walborn \&
  Liller}{1975}]{WalbornLiller} Walborn, N.R. \& Liller, M.H.\ 1977,
  ApJ, 211, 181

\bibitem[\protect\citeauthoryear{Walborn et al.}{Walborn et
  al.}{2002a}]{Walborn02a} Walborn, N.R., Danks, A.C., Vieira, G., \&
  Landsman, W.B.,\ 2002a, ApJS, 140, 407

\bibitem[\protect\citeauthoryear{Walborn et al.}{Walborn et
  al.}{2002b}]{Walborn02b} Walborn, N.R., Howarth, I.D., Lennon, D.J.
  et al.\ 2002b, AJ, 123, 2754

\bibitem[\protect\citeauthoryear{Walborn et al.}{Walborn et
  al.}{2007}]{Walborn07} Walborn, N.R., Smith, N., Howarth, I.D.,
  Vieira, G. et al.\ 2007, PASP, 119, 156

\bibitem[\protect\citeauthoryear{Walsh}{Walsh}{1984}]{Walsh84} Walsh,
 J.R.\ 1984, A\&A, 138, 380

\bibitem[\protect\citeauthoryear{Welch et al.}{Welch et
    al.}{1987}]{Welch87} Welch, W.J., Dreher, J.W., Jackson, J.M. et
    al.\ 1987, Science, 238, 1550

\bibitem[\protect\citeauthoryear{Whiteoak \& Otrupcek}{Whiteoak \&
  Otrupcek}{1984}]{Whiteoak84} Whiteoak J.B. \& Otrupcek R.E.\
  1984, PASA, 5(4), 552

\bibitem[\protect\citeauthoryear{Whiteoak}{Whiteoak}{1994}]{Whiteoak94}
Whiteoak J.B.Z.\ 1994, ApJ, 429, 225

\bibitem[\protect\citeauthoryear{Whitney et al.}{Whitney et
al.}{2004}]{whitney04} Whitney, B., Indebetouw, R., babler, B.L. et
al.\ 2004, ApJS, 154, 315

\bibitem[\protect\citeauthoryear{Yonekura et~al.}{Yonekura
  et~al.}{2005}]{Yonekura05} Yonekura Y., Asayama S., Kimura K.
  et~al.\ 2005, ApJ, 634, 476

\bibitem[\protect\citeauthoryear{Zhang et~al.}{Zhang
et~al.}{2001}]{Zhang01} Zhang X., Lee Y., Bolatto A., \& Stark A.A.\
2001, ApJ, 553, 274

\end{thebibliography}
\bibliographystyle{apj}

\end{document}